\documentclass[%
 reprint,
 superscriptaddress,
 nofootinbib,
nobibnotes,
 amsmath,amssymb,
 aps,
 prl,
]{revtex4-2}

\usepackage{graphicx}
\usepackage{dcolumn}
\usepackage{bm}
\usepackage[colorlinks,linkcolor=blue,urlcolor=blue, citecolor=blue]{hyperref}
\usepackage{lipsum}

\usepackage{dsfont}
\usepackage{physics}
\usepackage{enumitem}
\usepackage{tcolorbox}
\usepackage{needspace}

\usepackage{amsthm} 
\usepackage{thmtools} 
\usepackage{thm-restate}

\declaretheorem{corollary,lemma,proposition,definition}

\usepackage{amsmath, amssymb, amsfonts, amsthm}

\usepackage[normalem]{ulem}

\newcommand{\E}{{\rm E}}
\newcommand{\Var}{{\rm Var}}

\usepackage[page]{appendix}

\renewcommand{\appendixtocname}{List of appendices}

\makeatletter
\let\oldappendix\appendices

\g@addto@macro\tableofcontents{%
  \let\tf@toc@orig\tf@toc
}
\renewcommand{\appendices}{%
  \renewcommand{\thesection}{}
  \let\tf@toc\tf@app
  \addtocontents{app}{\protect\setcounter{tocdepth}{1}}
  \immediate\write\@auxout{%
    \string\let\string\tf@toc\string\tf@app
  }
  \oldappendix
}%

\g@addto@macro\endappendices{%
  \let\tf@toc\tf@toc@orig
  \immediate\write\@auxout{%
    \string\let\string\tf@toc\string\tf@toc@orig
  }%
}  

\renewcommand\tableofcontents{%
    \@starttoc{toc}%
}

\newcommand{\listofappendices}{%
  \begingroup
  \newcommand{\contentsname}{\appendixtocname}
  \let\@oldstarttoc\@starttoc
  \def\@starttoc##1{\@oldstarttoc{app}}
  \tableofcontents
  \endgroup
}

\makeatother

\begin{document}

\preprint{APS/123-QED}

\title{Entropic costs of extracting classical ticks from a quantum clock}

\author{Vivek Wadhia}
\thanks{Equal contributions}
\email[]{vivek.wadhia@eng.ox.ac.uk}
\affiliation{Department of Engineering Science, University of Oxford, Parks Road, Oxford OX1 3PJ, United Kingdom}

\author{Florian Meier}
\thanks{Equal contributions}
\email[]{florianmeier256@gmail.com}
\affiliation{Atominstitut, Technische Universität Wien, 1020 Vienna, Austria}

\author{Federico Fedele}
\affiliation{Department of Engineering Science, University of Oxford, Parks Road, Oxford OX1 3PJ, United Kingdom}

\author{Ralph Silva}
\affiliation{Institute for Theoretical Physics, ETH Z\"urich, Wolfgang-Pauli-Strasse 27, 8093 Z\"urich, Switzerland}

\author{Nuriya Nurgalieva}
\affiliation{Department of Physics, University of Zurich, Winterthurerstrasse 190, 8057 Zurich, Switzerland}

\author{David L. Craig}
\affiliation{Department of Materials, University of Oxford,
Oxford OX1 3PH, United Kingdom}

\author{Daniel Jirovec}
\affiliation{Institute of Science and Technology Austria, Am Campus 1, 3400 Klosterneuburg, Austria}

\author{Jaime Saez-Mollejo}
\affiliation{Institute of Science and Technology Austria, Am Campus 1, 3400 Klosterneuburg, Austria}

\author{Andrea Ballabio}
\affiliation{L-NESS, Physics Department, Politecnico di Milano, via Anzani 42, 22100, Como, Italy}

\author{Daniel Chrastina}
\affiliation{L-NESS, Physics Department, Politecnico di Milano, via Anzani 42, 22100, Como, Italy}

\author{Giovanni Isella}
\affiliation{L-NESS, Physics Department, Politecnico di Milano, via Anzani 42, 22100, Como, Italy}

\author{Marcus Huber}
\affiliation{Atominstitut, Technische Universität Wien, 1020 Vienna, Austria}
\affiliation{IQOQI Vienna, Austrian Academy of Sciences, Boltzmanngasse 3, 1090 Vienna, Austria}

\author{Mark T. Mitchison}
\affiliation{School of Physics, Trinity College Dublin, College Green, Dublin 2, D02 K8N4, Ireland}
\affiliation{Department of Physics, King’s College London, Strand, London, WC2R 2LS, United Kingdom}


\author{Paul Erker}
\affiliation{Atominstitut, Technische Universität Wien, 1020 Vienna, Austria}
\affiliation{IQOQI Vienna, Austrian Academy of Sciences, Boltzmanngasse 3, 1090 Vienna, Austria}

\author{Natalia Ares}
\email[]{natalia.ares@eng.ox.ac.uk}
\affiliation{Department of Engineering Science, University of Oxford, Parks Road, Oxford OX1 3PJ, United Kingdom}

\date{\today}

\begin{abstract}
We experimentally realize a quantum clock by using a charge sensor to count charges tunneling through a double quantum dot (DQD). Individual tunneling events are used as the clock's ticks. We quantify the clock's precision while measuring the power dissipated by the DQD and, separately, the charge sensor in both direct-current and radio-frequency readout modes. This allows us to probe the thermodynamic cost of creating ticks microscopically and recording them macroscopically. Our experiment is the first to explore the interplay between the entropy produced by a microscopic clockwork and its macroscopic measurement apparatus. We show that the latter contribution not only dwarfs the former but also unlocks greatly increased precision, because the measurement record can be exploited to optimally estimate time even when the DQD is at equilibrium. Our results suggest that the entropy produced by the amplification and measurement of a clock's ticks, which has often been ignored in the literature, is the most important and fundamental thermodynamic cost of timekeeping at the quantum scale.

\end{abstract}

\maketitle

\begin{figure*}
    \centering
    \includegraphics[width=\textwidth]
    {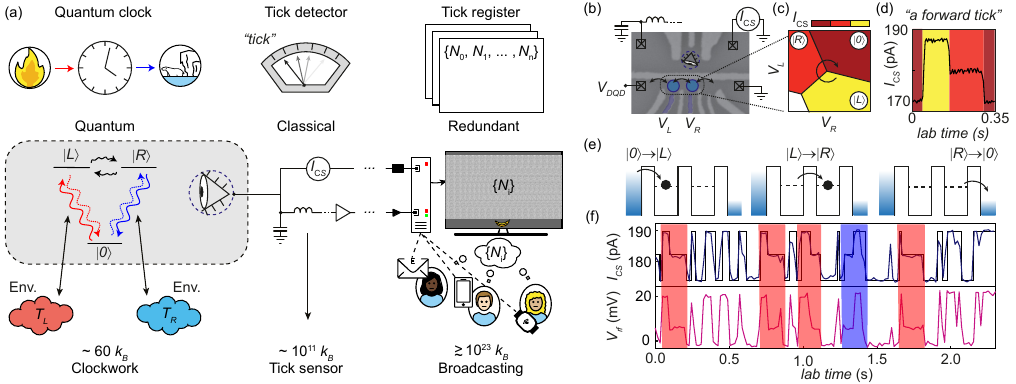}
    \caption{Concept versus experiment. 
    (a) A DQD with charge occupation states $\ket{0},\ket{L},\ket{R}$ stochastically exchanging charges with thermal environments can be used as a clock by identifying quantum jumps as \textit{ticks} (experimentally $\sim 60 k_B$ per tick, see value estimated as a maximum from Fig.~\ref{fig:panel2}(a)).
    Classically recording ticks to make them redundantly accessible for macroscopic agents requires measuring the quantum state and turning it into redundant information (right side).
    We found the cost of extraction to lie between $\sim 10^{9}\,k_B$ and $\sim 10^{11} k_B$ per tick in our experiment, depending on the measurement method, see respective values shown in Fig.~\ref{fig:panel2}(b,c).
    Nevertheless readout dissipation is several orders of magnitude smaller than macroscopic dissipation of the order $\gtrsim 10^{23}k_B$.
    (b) Scanning electron microscope image of a device similar to the one used in this experiment. Blue circles represent a DQD and the eye represents a charge sensor dot used for classical readout. $V_{L}$ and $V_{R}$ are the voltages applied to gate electrodes that control the number of charges within the left and right dot respectively. Charge transport can occur from either source to drain, drain to source, or any intermediary process (black arced arrows), influenced by an applied source-drain voltage $V_{\rm DQD}$ across the DQD.
    (c) Sketch of a charge stability diagram. Intersection of charge occupation states $\ket{0}$, $\ket{L}$ and $\ket{R}$ is a triple point.
    Fluctuations about this point cause a cycling of states (black arrow) which correspond to a forward clock tick.
    (d) The simplest case of a forward clock tick measured with current in relation to lab time. The trace forms a discrete, three-level telegraph signal that corresponds to the charge occupation states in (c). 
    (e) Schematic representation of the same forward clock tick as (d) in terms of charge transport. Initially, a charge enters from the source to the left dot $\ket{0} \rightarrow \ket{L}$, moves to the right dot $\ket{L} \rightarrow \ket{R}$, and finally moves to the drain $\ket{R} \rightarrow \ket{0}$.
    (f) Readout trace from both dc (top) and rf reflectometry (bottom), with respect to lab time. Black trace highlights the result of our level-identification algorithm.
    Shaded regions indicate ticks:
    red for forward and blue for backward. 
    }
    \label{fig:panel}
\end{figure*}

Clocks are physical systems tracking the passage of time, usually by quantifying time in discrete units called `ticks' which are stored in a register.
In practice this may, for example, be the clock's hand counting pendulum oscillations~\cite{Milburn2020,Pietzonka2022}, a digital counter counting laser oscillations in an atomic clock~\cite{Ludlow2015,Nicholson2015,Bothwell2022}, or even the amount of extant carbon-14 in dead organic material counting decay events~\cite{Milburn2020}.
One property all clocks have in common is that they rely on entropy increase due to the second law of thermodynamics to create an irreversible record of how much time has passed~\cite{Milburn2020,Barato2016,Erker2017,Pearson2021,Schwarzhans2021,Dost2023, Pietzonka2024}.
For all practical purposes, a clock's counter thus always increases, which aligns with our experience that clocks always tick forward. In the microscopic realm, however, quantum phenomena generally produce orders of magnitude less entropy than classical processes, leading to stochastic fluctuations~\cite{Crooks1999} that may cause a clock to tick backward on the quantum scale~\cite{Silva2023}.

This behavior is exemplified in Fig.~\ref{fig:panel}(a): a three-state system $(0,L,R)$ is the quantum clockwork (left), measurement apparatus (center), and the macroscopic agent (right).
While the example is presented with incoherent semi-classical dynamics, the conclusions we draw also hold for an adaptation to coherent dynamics, see details in SM~\cite{supp}.
In this example, the transition $0 \rightarrow L$ is driven by a hot thermal bath, $L \leftrightarrow R$ is balanced bidirectionally and does not rely on an external environment, $R \rightarrow 0$ is mediated by a cold thermal bath. Overall, the external environments induce a cycle $0 \rightarrow L \rightarrow R \rightarrow 0$ that transfers heat from hot to cold, producing entropy in accordance with the second law of thermodynamics. Considering these elementary cycles as ticks of the clock, the hot-cold bias drives the clock forward. Nevertheless, stochastic fluctuations may sometimes transfer heat against the thermal gradient, causing the microscopic clock to tick backward. The extreme case being both baths are brought to the same temperature, where both tick directions are equally likely, no entropy is produced on average, and the clock stops ticking.

By continuously monitoring the microscopic clockwork, however, an observer could register each stochastic transition ($0\to L$, $L\to 0$, etc.) as a new tick. This would restore the unidirectional flow of time recorded by the clock, even in the absence of a thermal gradient driving the clockwork. Yet a paradox arises: a clock is a manifestly irreversible device that distinguishes past from future --- how can it produce no entropy? The resolution, as we verify in this work, is the measurement process itself produces entropy while creating a classical record of the quantum clock's ticks~\cite{Granger2011,Jacobs2012,Deffner2016,Debarba2019,Guryanova2020,Chu2022, *Chu2022b, Dooley2022, Schwarzhans2023,Latune2025}.


To demonstrate this, we experimentally realize the quantum-clock model of Fig. 1(a) using a double quantum dot (DQD). Ticking of this clock is probed via direct current (dc) and radio-frequency (rf) reflectometry readout modes, whose power dissipation can be directly inferred. We are thus able to measure two distinct sources of entropy production driving the clock: (1) the entropy produced by the quantized clockwork, and (2) the entropic cost of generating a classical measurement record. The first contribution has been subject of several studies showing more regular ticks generally require more dissipation~\cite{Barato2016,Erker2017,Pearson2021,Dost2023,Schwarzhans2023,Silva2023,Pietzonka2024}.
This principle is reminiscent of more general thermodynamic uncertainty relations (TURs). For classical stochastic systems~\cite{Barato2015,Gingrich2016,Pietzonka2017,Macieszczak2018,Horowitz2020} and quantum systems~\cite{Ptaszyifmmodenelsenfiski2018,Agarwalla2018,Liu2019a,Guarnieri2019,RignonBret2021,Prech2023,Falasco2020,Hasegawa2021}, TURs bound microscopic fluctuations by entropy production, though in certain settings the bounds are quite loose~\cite{Pietzonka2022,Brandner2018,Gerry2022}.
Nevertheless, our measurements show that the second contribution --- the entropic cost of extracting classical ticks from the quantum clock --- is, in fact, the dominant one by 9 orders of magnitude, highlighting the separation of scales between microscopic irreversibility and irreversibility induced by the measurement.
We also show that exploiting all information in the measurement record to optimally estimate time~\cite{Prech2024}, boosts the clock's precision by an order of magnitude compared to naively counting charge-transfer cycles.

\paragraph*{Experimental realization.}
We realize a quantum clock using charge occupation states of a DQD. Charge transport through the DQD constitutes the ticking mechanism of the clock.
A charge sensor dot, capacitively coupled to the DQD, is used to monitor the DQD, converting its charge occupation state into a classical signal~\cite{Charge_sensor, Elzerman2003}. The quantum dots are defined electrostatically using a Ge/SiGe heterostructure with gate electrodes patterned on top \cite{Lateral_qd}, similar to Fig.~\ref{fig:panel}(b). The semiconductor device was loaded into a dilution refrigerator and cooled down to $T=180$ mK.
Typical charging energies for these devices are $\sim$ 1.3 meV.


Transport of a charge between source and drain defines clock ticks. This process can be divided into three states, a charge in the left dot $\ket{L}$, a charge in the right dot $\ket{R}$, or no charge in either dot $\ket{0}$.  A \textit{forward} tick is any sequence that starts with $\ket{0} \rightarrow \ket{L}$ and ends with $\ket{R} \rightarrow \ket{0}$ with no $\ket{0}$ between. For example, $\ket{0} \rightarrow \ket{L} \rightarrow \ket{R} \rightarrow \ket{0}$ like in Fig.~1(e), but also $\ket{0} \rightarrow \ket{L} \rightarrow \ket{R} \rightarrow \ket{L} \rightarrow \ket{R} \rightarrow \ket{0}$ etc. A \textit{backward} tick, conversely, is defined as starting with $\ket{0} \rightarrow \ket{R}$ and ending with $\ket{L} \rightarrow \ket{0}$ with no $\ket{0}$ between.
Superpositions of DQD charge states are not considered as part of this experiment because they decohere faster than the measurement rate.
Thus, the DQD is always in a definite charge occupation state when measured, which is not substantially impacted by the measurement.
By applying a voltage bias across the double dot $V_{\rm DQD}$, we effectively have two thermal baths (given by source and drain reservoirs) out of equilibrium, and one of these tick processes becomes favored over the other. Each tick results in the dissipation of energy $q=e|V_{\rm DQD}|$, where $e$ is the elementary charge. For environments at inverse temperature $\beta$, this equals an entropic cost per tick
\begin{equation}
\label{eq:SigmaTick}
    \Sigma_{\rm tick}=\beta e|V_{\rm DQD}|.
\end{equation}
	
In this experiment, we use two distinct methods to record ticks. Firstly, by measuring changes in the current flowing through the charge sensor dot (dc charge sensing).
Secondly, through radio-frequency (rf) reflectometry that measures changes in the sensor's reflectance given a fixed rf input signal.
Applying a voltage bias $V_{\rm cs}$ to the charge sensor generates a current $I_{\rm cs}$, that varies with the charge occupation of the DQD.
Fig.~\ref{fig:panel}(d) shows an example of a dc measurement: three distinct current levels appear that correspond to $\ket{0}$, $\ket{R}$ and $\ket{L}$. Creating a classical record of the DQD state thus comes at a fundamental entropic cost due to the power dissipated through the sensor dot, given by Ohm's law $P^{\rm dc}_{\rm diss} = V_{\rm cs} I_{\rm cs}$. In order to lower the dissipation $V_{\rm cs}$ must be decreased. However this results in a decrease in $I_{\rm cs}$, which in turn decreases the signal-to-noise ratio (SNR). Qualitatively a lower SNR equates to less difference between the three levels of the telegraphic signal. Sec.~\ref{supp:low_readout_regime} of the SM~\cite{supp} provides a more detailed explanation.
	
For rf reflectometry, a reference signal with a fixed power $P_{\rm in}$ is sent to an rf cavity coupled to the charge sensor. It is then reflected back carrying information about the charge occupation state of the DQD. Using demodulation, traces like Fig.~\ref{fig:panel}(f) can be obtained. The change in the reflected signal results in a reflected power $P_{\rm out}$, that is smaller in magnitude than $P_{\rm in}$. By accounting for known losses that occur in the signal chain, it is possible to compare these two quantities and establish the power dissipated across the sensor when using reflectometry. A more detailed description of this calculation is given in Sec.~\ref{supp:rf_power_diss_calc} in the SM~\cite{supp}. To modulate the SNR, the input signal frequency $f_{\rm rf}$ was gradually increased, moving it progressively off-resonance with the rf cavity. As with dc measurements, this leads to a change in the power dissipated and hence the SNR.   
	
\paragraph*{Experimental protocol.}
$V_{\rm DQD}$ was initially set to 0~mV to simulate two baths in thermal equilibrium. $V_{\rm cs}$ was set to 0.345~mV. An rf signal was applied to the charge sensor, with frequency $f_{\rm rf}= 114$~MHz and power $P_{\rm rf}= -20$~dBm. These initial settings were chosen to optimize the SNR of the sensor signal. The acquisition card's integration time, both for the dc and rf sensor configurations, was set to 5~ms. A suitable triple point in the charge stability diagram was then selected such that transitions between states $\ket{0}$, $\ket{L}$, and $\ket{R}$ occur purely due to stochastic fluctuations. That is, the system was not driven by an external ac signal. Following this, transitions between these three states were slowed down until they appeared as a three-level telegraphic signal as in Fig.~\ref{fig:panel}(f). This was achieved by gradually increasing the gate voltages controlling the tunnel barriers of the DQD, and estimating the tunneling rates (also referred to as jump rates) between the dots until they were all equal. Once suitable rates were achieved, these gate voltages remained fixed for the remainder of the experiment.

A 30-minute measurement was then taken of the current through the charge sensor $I_{\rm cs}(t)$, and the reflected rf signal $V^{\rm rf}_{\rm out}(t)$. $I_{\rm cs}(t)$ and $V^{\rm rf}_{\rm out}(t)$ were filtered and amplified, with $V^{\rm rf}_{\rm out}(t)$ also demodulated, in order to obtain classical signals for readout. These signals were then used to identify clock ticks, as detailed in SM~\cite{supp} Sec.~\ref{supp:details_experiment}. From this data, it is possible to estimate the entropic cost of the clock ticking. In order to estimate and compare the entropic cost of the measurement process, the SNR needs to be modulated. This was achieved by decreasing $V_{\rm cs}$ from 0.345~mV to 0.102~mV whilst increasing $f_{\rm rf}$ from 114~MHz to 128~MHz, in four steps. In other words, there are four settings of the SNR ranging from very good to very poor. The process of recording 30-minute measurements was then repeated for each SNR setting. Finally, to simulate the effect of thermal baths out of equilibrium, $V_{\rm DQD}$ was changed from $0$ to $+1$~mV for nine settings, with four SNR measurements taken at each value of $V_{\rm DQD}$.

\begin{figure*}
    \centering
    \includegraphics[width=\textwidth]{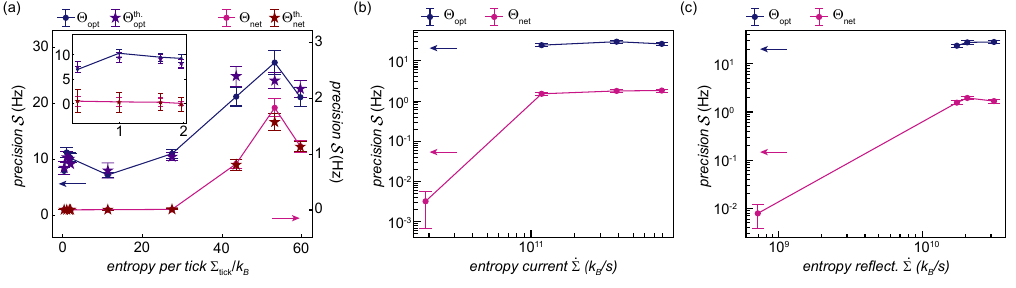}
    \caption{Clock precision and entropy trade-off (empirical data and theory comparison).
    (a) Here, the precision $\mathcal S$ of the clock defined by the net number of charges $\Theta_{\rm net} = \nu^{-1} N(t)$ transported across the DQD is plotted against the entropy dissipated per tick (scale on the right).
    Furthermore, the normalized precision of the optimal estimator $\Theta_{\rm opt}$ from eq.~\eqref{eq:Theta[i]} is shown, also as a function of entropy per charge transport (scale on the left).
    The dots indicate empirical estimates for the precision (SM~\cite{supp} Sec.~\ref{supp:empirical_precision}), and stars the theory prediction from the Markovian model (SM~\cite{supp} Sec.~\ref{supp:theoretical_precision}).
    Drifts in the device can lead to changes in the transition rates and ultimately deviations between theoretical and empirical values.
    (b) Data obtained with the dc measurement of the sensor dot showing how precision $\mathcal S$ 
    for $\Theta_{\rm net}$ and $\Theta_{\rm opt}$ scales as a function of the dissipation rate in the sensor dot.
    (c) Data and dissipation for the reflectometry method, exhibiting three orders of magnitude lower dissipation than the dc method.
    Error bars show the estimated standard error as calculated in the SM~\cite{supp}.}
    \label{fig:panel2}
\end{figure*}

\paragraph*{Theoretical description.}

The present experiment can be described by a classical master equation~\cite{Cuetara2015,Bettmann2023},
\begin{align}
\label{eq:dotp=Mp}
    \dot{\mathbf p} = M \mathbf p,
\end{align}
where $\mathbf p = (p_0,p_R,p_L)$ denotes charge state occupation probability, and $M$ the transition rate matrix.
The validity of this model is discussed in Sec.~\ref{supp:characterizing_rates} of the SM~\cite{supp}.
Each experimental run yields a stochastic trajectory~\cite{VandenBroeck2015,Carmichael1993} of the master equation~\eqref{eq:dotp=Mp}.
In the readout signal $I_{\rm cs}(t)$, from either the dc or rf methods, the states $\ket{0}$, $\ket{R}$ and $\ket{L}$ manifest as three signal values $I_{\rm l}<I_{\rm m}<I_{\rm h}$.
Due to noise, the signal does not exactly take one of the three values, moreover the three values are a priori unknown.
Initially, the readout has to be calibrated by measuring the $I_{\rm cs}(t)$ for some time and creating a histogram of the measured values.
For sufficiently high readout SNR, the histogram exhibits three distinct peaks centered on the values $I_{\rm l/m/h}$, where each peak is identified with one of the three microscopic states with an estimated identification error between 0.1\% and 1\% (see Fig.~\ref{fig:ID_error} and Sec.~\ref{supp:tick_readout} in the SM~\cite{supp} for details).
Through this identification, the measured signal $I_{\rm cs}(t)$ is converted into a sequence $S(t)\in\{0,R,L\}$ of states as shown in Fig.~\ref{fig:panel}(f).
While the signal analyzer includes a clock to assign time tags $t$ to the measured states, this is only required for calibration.
To use the device as a clock, determining $t$ is the goal.
Thus, only the instantaneous state $S(t)$ is required and not $t$.
This is achieved by recording the sequence of charge occupation states occupied up to `$t$', starting with the initial state $s_0$, the state after the first jump $s_1$ and so on. We denote this sequence as $\mathbf s(t) = (s_0,s_1,\dots,s_{m(t)})$, where $m(t)$ the number of jumps recorded.

\paragraph*{Time estimation.}
The time estimator $\Theta_{\rm net}$ can be reconstructed based on the record $\mathbf s(t)$ by counting all the forward ticks as $+1$ and all the backward ticks as $-1$ towards the net number of transfers $N[\mathbf s(t)]$.
Finally, $\Theta_{\rm net}[\mathbf s(t)] = \nu^{-1}N[\mathbf s(t)]$ is obtained by normalizing by the average charge transfer rate $\nu$, which can be initially calibrated to express $\Theta_{\rm net}$ in some reference unit, e.g.~seconds.
The uncertainty of the calibrated frequency can be made arbitrarily small by increasing the calibration time (Sec.~\ref{supp:empirical_precision} of the SM~\cite{supp}).

While $\Theta_{\rm net}$ is how the ticks are naturally encoded in the charge environments, it is a suboptimal way of estimating time after readout where the full jump sequence $\mathbf s(t)$ is (classically) available.
This becomes particularly apparent when considering the equilibrium case $V_{\rm DQD}\approx 0\,{\rm mV}$,
where no net charge transfer takes place $N[\mathbf s(t)]\approx 0$ and the passage of time is not recorded microscopically.
For stochastic evolution following eq.~\eqref{eq:dotp=Mp}, a theoretically optimal unbiased time estimator linear in the number of observed jumps exists~\cite{Prech2024}.
Unbiased refers to the expectation value $\E[\Theta(t)]=t$ being equal to the lab time $t$, optimality to the precision defined in eq.~\eqref{eq:S}.
This optimal choice is
\begin{align}
\label{eq:Theta[i]}
    \Theta_{\rm opt}[\mathbf s(t)] = \sum_{k=0}^{m(t)} \frac{1}{\Gamma_{s_k}} = \sum_{s\in\{0,R,L\}} \frac{n_s(t)}{\Gamma_s},
\end{align}
$n_s(t)$ is the number of times the state $s$ appears in the sequence $\mathbf s(t)$.
The sum weights each jump from a state $s$ by the average time it takes for the system to leave the state, which is given by $1/\Gamma_{s} = -1/M_{ss}$.
Similar to the normalization $\nu$ for $\Theta_{\rm net}$, the rates $\Gamma_s$ can be initially calibrated with respect to a lab clock.

One way of quantifying the quality of the time estimate $\Theta$ is to compare the expectation value $\E[\Theta(t)]$ with the variance $\Var[\Theta(t)]$.
We consider the precision
\begin{align}
\label{eq:S}
    \mathcal S = \lim_{t\rightarrow\infty} \frac{\E[\Theta(t)]^2}{t\, \Var[\Theta(t)]},
\end{align}
which captures the absolute timescale at which the estimator fluctuates in units of a rate, e.g.~Hz.
The larger the precision $\mathcal S$, the faster the process that can be estimated reliably with the clock (details in the End Matter).
SM~\cite{supp} includes pertinent detail how $\mathcal S$ is obtained empirically (Sec.~\ref{supp:empirical_precision},) and theoretically (Sec.~\ref{supp:theoretical_precision}).

\paragraph*{Thermodynamic analysis.}
In Fig.~\ref{fig:panel2}, precision $\mathcal S$ is plotted against the dissipation in the DQD in (a), the sensor dot for dc measurement in (b) and rf measurement in (c).
The comparison in Fig.~\ref{fig:panel2}(a) reveals a trade-off between the dissipation $\Sigma_{\rm tick}$ in the clockwork and the precision of $\Theta_{\rm net}$.
As expected, the precision of the microscopic time record vanishes when the DQD is brought to equilibrium.
This is also in agreement with the TUR~\cite{Barato2015,Gingrich2016,Pietzonka2017,Macieszczak2018,Horowitz2020} which translates to $\mathcal S\leq \nu \Sigma_{\rm tick}/2$.
Nevertheless, the optimal estimator $\Theta_{\rm opt}$ still works even when the clockwork is at equilibrium.
Ultimately this is possible because of the dissipation required for the measurement to create the irreversible record of the equilibrium DQD dynamics.
When the readout SNR is sufficiently high, as shown in Fig.~\ref{fig:panel2}(b,c), both $\Theta_{\rm net}$ and $\Theta_{\rm opt}$ yield high precision.
However, when the measurement dissipation drops below a method dependent threshold, both time estimates become ill-defined and the clock stops ticking altogether.

\paragraph*{Discussion.}
It is well established that quantum clocks produce entropy to generate the ticks necessary to measure time~\cite{Milburn2020,Barato2016,Erker2017,Pearson2021,Schwarzhans2021,Dost2023, Pietzonka2024}.
However, the critical role of measurement irreversibility in reading the clock's ticks has remained unexplored. Although a recent experiment has investigated measurement backaction on a quantum clock~\cite{He2023}, our work is the first experiment probing the thermodynamic cost of this step. In doing so, we have demonstrated charge transport across a DQD can be used to construct a microscopic quantum clock, which is a key step towards building autonomous quantum machines~\cite{Meier2024,Woods2024,Cilluffo2024,MarinGuzman2024}.

In our experiment, we observed rf reflectometry is thermodynamically more efficient than dc measurement. For both methods, however, we found that dissipation is fundamentally linked to the SNR of the readout signal, and a minimum amount of dissipation is needed to create a classical record of the quantum ticks. Whilst we confirmed the clock's precision is improved by increasing entropy production in the clockwork, in accordance with previous work~\cite{Milburn2020,Barato2016,Erker2017,Pearson2021,Schwarzhans2021,Dost2023, Pietzonka2024}, this dissipation is eclipsed by the thermodynamic cost of the measurement. We have shown that the latter is particularly important for timekeeping because it can drive the clock's operation, even in scenarios where the clockwork itself ceases to generate entropy. 
While the model studied in this work is based on a DQD setup, extracting timing information from a quantum system generally comes at a thermodynamic cost, including in atomic or optical setups which currently form the basis of state-of-the-art clocks~\cite{Ludlow2015,Nicholson2015,Bothwell2022}.
By establishing that tick extraction generally dominates the entropic cost of timekeeping, our work may guide future microscopic clock designs to improve their precision in the most thermodynamically efficient way.

\section*{Acknowledgements}
The authors thank Georgios Katsaros for providing the device for this experiment, 
and Tony Apollaro, Ilia Khomchenko and Gerard Milburn for discussions.
VW acknowledges funding from UK Research and Innovation grant EP/T517811/1.
FM, MH and PE acknowledge funding from the European Research Council (Consolidator grant ‘Cocoquest’ 101043705).
MH and PE acknowledge funding from the Austrian Federal Ministry of Education, Science, and Research via the Austrian Research Promotion Agency (FFG) through Quantum Austria.
RS acknowledges funding from the Swiss National Science Foundation via an Ambizione grant PZ00P2\_185986.
MTM is supported by a Royal Society University Research Fellowship.
NA acknowledges support from the European Research Council (grant agreement 948932) and the Royal Society (URF-R1-191150). 
This project is co-funded by the European Union (Quantum Flagship project ASPECTS, Grant Agreement No.\ 101080167) and UK Research and Innovation (UKRI). Views and opinions expressed are however those of the authors only and do not necessarily reflect those of the European Union, Research Executive Agency or UKRI. Neither the European Union nor UKRI can be held responsible for them.

\section*{Data availability}
The measurement traces as well as the code for analysis is publicly available on GitHub~\cite{Git2025}.

\section*{Author contributions}
FM analyzed the data and developed the theory framework with inputs from RS, NN, MTM and PE.
VW performed the experiment supervised by FF and with input from DLC.
DJ and JSM fabricated the device, AB, DC and GI provided the heterostructure.
FM drafted the initial version of the manuscript.
VW wrote the experimental sections.
MH conceptualized the narrative.
MTM, PE and NA supervised the project.
MTM, FM, FF and NA conceived the original idea of the experiment. 
All authors contributed to the discussions of the results and finalization of the manuscript.

\bibliography{bibliography}

\clearpage\newpage
\onecolumngrid
\section*{End Matter}

\begin{figure*}
    \centering
    \includegraphics[width=\linewidth]{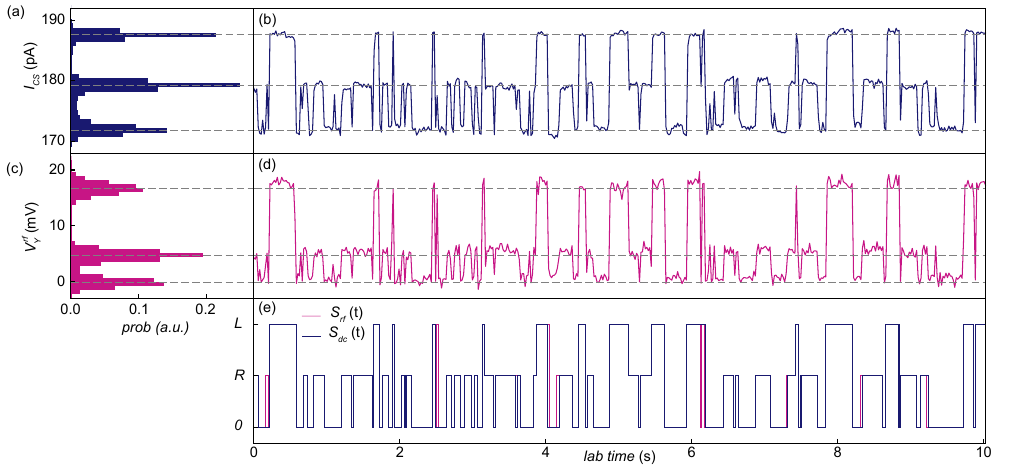}
    \caption{(Extended Data) Measurement traces of length $10\,{\rm s}$ using both the dc (b) and the rf reflectometry method (d).
    The trace is taken for a DQD bias $V_{\rm DQD}=0\,{\rm mV}$, a sensor dot bias $V_{\rm cs}=0.345\,{\rm mV}$, and frequency $f_{\rm rf}=114\,{\rm MHz}$ for the rf reflectometry drive.
    (a) Histogram $p(I)$ of the current values for the full $30\,{\rm min}$ data trace showing the emergence of three peaks corresponding to the DQD charge occupation states, shown as the dashed lines.
    (c) Analogously, the histogram $p(V)$ for the demodulated Y-component of rf sensor signal is shown here (details in Sec.~\ref{supp:details_reflectometry} of the SM~\cite{supp}).
    (e) The discretized signal $S(t)$ for both the dc and rf reflectometry methods, obtained from the measurement traces by identifying one of the three states from the noisy signal.
    For most times, the discretized signals from the two readout methods agree with each other.
    In general, rf is more sensitive than dc sensing, leading to instances where the two discretized signals are not exactly identical. Nevertheless, as can be seen, this difference does not significantly affect level identification.
    }
    \label{fig:endmatter}
\end{figure*}

\twocolumngrid

\paragraph*{Tick identification method.}
Here, we provide pertinent detail regarding our method to identify ticks from the measurement signal.
To create a classical record of the DQD clock's ticks, the two measurement signals obtained through either the dc or rf sensor method have to be post-processed to infer the state of the DQD.
In Fig.~\ref{fig:endmatter}, we show a comparison of a current signal $I_{\rm cs}(t)$ (dc), with the rf sensor voltage $V_{\rm Y}^{\rm rf}(t)$ (rf).
The signal $V_{\rm Y}^{\rm rf}$ is the Y quadrature component of the rf signal measured with an rf lock-in amplifier after demodulation, chosen for illustration purposes (details in Sec.~\ref{supp:details_reflectometry} of SM~\cite{supp}).
From comparison of Figs.~\ref{fig:endmatter}(b) versus Fig.~\ref{fig:endmatter}(d), it can be seen that both readout methods yield qualitatively similar outputs, up to a difference in noise characteristics.
Still, both outputs can then be used to infer the the state of the DQD by identifying the three separated levels around which the signals fluctuate.
We use the following method to identify the three levels from the given signal (further details in Sec.~\ref{supp:identifying_states} of SM~\cite{supp}).

Initially, the output signals have to be calibrated to determine the three values $I_{\rm l}<I_{\rm m}<I_{\rm h}$.
In the following, we describe the identification method for the dc sensor readout, but the method works analogously for the voltage output from the rf reflectometry.
In principle, one way to calibrate the readout is to initially run the device for some time and create a histogram $p(I)$ for how likely a given output current value is.
From the histogram, three peaks then reveal the three current values corresponding to the three states of the DQD.
After the initial calibration, the output signal can then be converted in real time into a discrete output encoding the DQD state, from which the clock's ticks can then be inferred.

Instead of a real-time identification, we have analyzed the present experiment's output \textit{a posteriori}.
For each of the 30 min data traces, we create a histogram for the output current values, shown in Fig.~\ref{fig:endmatter}(a), or respectively voltage values shown in Fig.~\ref{fig:endmatter}(c).
Based on the method from~\cite{Craig2023}, we numerically fit three Gaussian peaks $\sum_{i=1}^3  h_i \mathcal N(\mu_i,\sigma_i)$ to the histogram, where $\mu_i$ is the mean and $\sigma_i$ the width of the $i$th Gaussian.
The prefactor $h_i$ are weights to account for the relative probability of each of the three levels.
From the fitted normal distributions, the measurement output can then be converted into the discrete signal shown in Fig.~\ref{fig:endmatter}(e), by mapping the current values $I_{\rm cs}(t)$ (or voltages $V^{\rm rf}(t)$) to the most likely of the three peaks from the histogram.

\paragraph*{Relating different notions of clock precision.}
In the following paragraphs we provide details for how different notions for clock precision in the literature are related to the definition of $\mathcal S$ in eq.~\eqref{eq:S}. There are three distinct ways to characterize a ticking clock's output.

\textit{1. Time between ticks.} If the clock's tick counter only increases, i.e., there are only forward ticks, one may consider the probability density $P[T=\tau]$ that the time between two subsequent ticks is $\tau\geq0$ (the \textit{waiting-time distribution}). In the literature~\cite{Erker2017,Woods2019,Woods2021,Woods2022,Schwarzhans2023,Dost2023,Nello2024} the \textit{accuracy} with which such a tick estimates the underlying parameter time is defined by 
\begin{align}
\label{eq:N_clockaccuracy}
    \mathcal N_\infty = \frac{\E[T]^2}{\Var[T]}.
\end{align}
If the ticks are independently and identically distributed (iid), making the clock a \textit{renewal process}, the accuracy $\mathcal N_\infty$ can be understood as the expected number of times the clock ticks until it goes wrong by one tick. It is thus the precision of the clock relative to its own timescale, and $\mathcal N_\infty$ is timescale invariant.

\textit{2. Counting of ticks.} When the number of ticks $N(t)$ is made up of forward ticks $N^+(t)$ and backward ticks $N^-(t)$ like for the DQD clock, the tick counter can sometimes jump backward or forward.
As a result, there is no longer a unique waiting time describing a strictly increasing counter.
By working in the framework of counting observables, one can thus consider the net number of ticks $N(t)$ instead of the waiting times.
In this case, one defines the precision of the counting observable $N(t)$, also known in the literature as the Fano factor~\cite{Barato2015,Barato2016},
\begin{align}
\label{eq:N_clockprecision}
    \mathcal N = \lim_{t\rightarrow\infty}\frac{\E[N(t)]}{\Var[N(t)]}.
\end{align}
This is well defined thanks to the following generalized relations~\cite{Prech2024,Landi2024},
\begin{align}
    \E[N(t)]=\nu t + O(1),\quad \Var[N(t)]=Dt + O(1),
\end{align}
with $\nu$ the tick rate and $D$ the diffusion constant. The Fano factor is time-invariant and is thus also a measure of relative precision.

If the clock is a renewal process (ticks are iid and only forward), then it is possible to relate the moments of the tick time to the moments of the tick number in the long-time limit~\cite{Cox1962,Silva2023},
\begin{align}
\label{eq:EN_VarN}
    E[T] = \frac{1}{\nu}, \quad &\Var[T] = \frac{D}{\nu^3}.
\end{align}
Thus, eqs.~\eqref{eq:EN_VarN} guarantee that for a renewal process the accuracy~\eqref{eq:N_clockaccuracy} in the waiting time picture equals the precision from eq.~\eqref{eq:N_clockprecision}, that is, $\mathcal N=\mathcal N_\infty$.

\textit{3. Time estimator from detailed jumps.} Finally, one may construct a linear time estimator $\Theta(t)$ from the number of each type of tick that the clock produces (details in SM~\cite{supp}, Sec.~\ref{supp:general_clock_theory}), which is assumed to be unbiased, i.e. $\E[\Theta(t)]=t + O(1)$.
Following~\cite{Prech2024}, we have defined the \textit{absolute precision} $\mathcal S$, given by eq.~\eqref{eq:S} in the main text and repeated here for convenience:
\begin{align}
\tag{\ref{eq:S} rev.}
    \mathcal S = \lim_{t\rightarrow\infty} \frac{\E[\Theta(t)]^2}{t\, \Var[\Theta(t)]}.
\end{align}
To understand the terminology of absolute precision, note that in the number-of-ticks picture one could define an unbiased time estimator as $\Theta_{\rm net}(t)=\nu^{-1} N(t)$, in which case the two notions of precision $\mathcal S$ and $\mathcal N$ are related by
\begin{align}
    \mathcal N = \frac{\mathcal S}{\nu}.
\end{align}
That is, the \textit{relative precision} $\mathcal N$ equals the absolute precision $\mathcal S$ normalized by the clock's resolution $\nu$.
This equality can be generalized even for time estimators such as the optimal estimator $\Theta_{\rm opt}(t)$ as discussed in Ref.~\cite{Prech2024}, though in that case, there is in general no well-defined integer number of ticks $N(t)$, because the different increments of $\Theta_{\rm opt}(t)$ are not necessarily integer multiples of each other.

In summary, the relative precision is the clock's precision with respect to its own timescale, while the absolute precision is the precision relative to a fundamental unit of time.

\clearpage
\newpage

\setcounter{secnumdepth}{2}
\appendix
\onecolumngrid
\section*{Supplemental Material}

\twocolumngrid

\listofappendices

\begin{appendices}
\section{\label{supp:details_experiment}Details on the experiment}

\subsection{\label{supp:details_current}dc charge sensing}

\begin{figure}[!h]
    \centering
    \includegraphics[width=\linewidth]{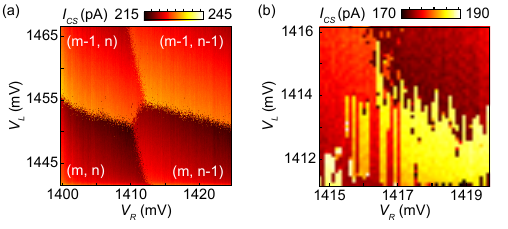}
    \caption{(a) Charge stability diagram. Colormap indicates the current flowing through the sensor dot as a function of $V_{L}$ and $V_{R}$. Indices m and n indicate the number of charges in the left and right dot, respectively. Note the charges are holes, not electrons. Boundaries of the honeycomb pattern correspond to the transitions between the charge occupation states of the DQD. The intersection of three boundaries indicates a triple point.
    (b) Closer view of a triple point in the latched regime.}
    \label{fig:cs_diagram}
\end{figure}

In order to measure the charge occupation of the DQD, the current flowing through the charge sensor dot $I_{\rm cs}$ was measured, initially with zero voltage bias across the DQD, $V_{\rm DQD} = 0$ mV and for optimal SNR, $V_{\rm cs} = 0.345$ mV. Measuring $I_{\rm cs}$ as a function of $V_{L}$ and $V_{R}$, produces a honeycomb-like pattern. This pattern is known as a charge stability diagram, and is shown in Fig.~\ref{fig:cs_diagram} (a). The edges of the honeycomb mark the transitions between charge occupation states. Considering that this device is operated in depletion mode, we can assign a relative charge occupation state based on this diagram. The charge stability diagram shows different values of $I_{\rm cs}$ for different regions of the honeycomb indicating different occupation states. Therefore there is a one-to-one correspondence between the current measured and the charge state of the DQD.

Due to the limited bandwidth of the dc sensor, in order to detect the charge transitions, tunneling rates must be slowed down. We thus tuned the DQD in the so called \emph{latched} regime, which is achieved by adjusting the tunnel barrier voltages of the DQD until all tunnelling events, including the interdot transitions, can be sensed. Fig.~\ref{fig:cs_diagram} shows the effect of latching on the charge stability diagram. The pixelation of the boundary between charge occupation states is characteristic of the latching regime.

From Fig.~\ref{fig:cs_diagram}, it can be seen how charge occupation states translate to the three states of the DQD. Explicitly, the state (m-1, n-1) = $\ket{0}$, (m, n-1) = $\ket{L}$ and (m-1, n) = $\ket{R}$.

\subsection{\label{supp:dc_power_diss_calc}Calculating power dissipation from dc charge sensing}
For dc charge sensing, the power dissipated is given by the product of the voltage drop across the sensor and the current flowing through it. In order to distinguish the three DQD states, the baseline current through the charge sensor is not relevant, since only the difference between the three levels matters.
We calculate the power dissipation (relative to the baseline current) as
\begin{align}
    \label{eq:power_diss_current}
    P^{\rm dc}_{\rm diss}(t) = V_{\rm cs}|I_{\rm cs}(t) - \langle I_{\rm cs}\rangle|,
\end{align}
where $I_{\rm cs}(t)$ is the instantaneous current at a given moment in time and $\langle I_{\rm cs}\rangle$ is the time averaged baseline current.
The voltage bias $V_{\rm cs}$ across the charge sensor remains fixed for a given trace.
In Fig.~\ref{fig:panel2}(b), the time-averaged value of $P^{\rm dc}_{\rm diss}(t)$ is then taken to determine the entropy dissipation at the dilution refrigerator's temperature, $T=180$~mK.

\subsection{\label{supp:details_reflectometry}rf charge sensing}

\begin{figure}[h]
    \centering
    \includegraphics{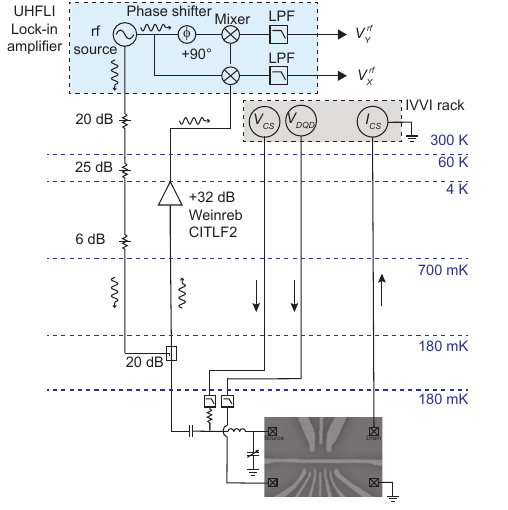}
    \caption{dc and rf sensor circuit. The device is mounted on a printed circuit board containing a lumped element rf cavity with an integrated bias tee. 
    Gray wavy lines show rf signal propagation and gray straight lines the dc current path. The rf input signal is generated by a Zurich Instruments UHFLI lock-in amplifier. The reflected signal is amplified at 4K 
    using a CITLF2 low-noise cryogenic amplifier. Dashed purple lines indicate the different stages of the dilution refrigerator. Dual-phase demodulation of the reflected rf signal occurs within the lock-in amplifier (blue shaded region). LPF stands for low pass filter.   
    }
    \label{fig:rf_signal_chain}
\end{figure}

Radiofrequency reflectometry is a widely used technique for quantum device readout~\cite{Vigneau2023}.
A fixed rf signal $V^{\rm rf}_{\rm in}(t)$ drives an rf cavity coupled to the charge sensor dot, with the reflected signal probing the changes in the sensor's impedance.
Changes in the DQD occupation, measured by the charge sensor, cause a change in the overall impedance of the device. The reflected signal $V^{\rm rf}_{\rm out}(t)$ therefore differs from $V^{\rm rf}_{\rm in}(t)$ in a way that captures the charge occupation state of the DQD. Specifically, the change in amplitude and phase of $V^{\rm rf}_{\rm out}(t)$ with respect to $V^{\rm rf}_{\rm in}(t)$ encodes this information. The reflected signal $V^{\rm rf}_{\rm out}(t)$ is amplified at 4K, and measured, demodulated and amplified at room temperature using an rf lock-in amplifier. Figure \ref{fig:rf_signal_chain} illustrates the setup for the rf sensor measurements. 

In order to obtain the best trace for tick identification, we combine the X and Y signal components of $V^{\rm rf}_{\rm out}(t)$, $V^{\rm rf}_{X}(t)$ and $V^{\rm rf}_{Y}(t)$, using the method of principal component analysis (PCA). This reconstructs one signal from quadrature components with minimal loss of information. A more detailed guide to this method is provided here \cite{Tharwat2021}. This final signal is referred to as simply $V_{\rm rf}(t)$.

\subsection{Calculating power dissipation from rf reflectometry}
\label{supp:rf_power_diss_calc}

The power dissipated by the rf sensor is the difference between the power input $P_{\rm in}$ and the power ouput $P_{\rm out}$:
\begin{equation}
    P^{\rm rf}_{\rm diss}(t) = P_{\rm in}(t)-P_{\rm out}(t).
\end{equation}
The power input $P_{\rm in}$ was fixed at -20 dBm at room temperature. The attenuation of this signal in the dilution refrigerator wiring is shown in Fig.~\ref{fig:rf_signal_chain}. $P_{\rm in}$ is thus -71 dBm. The output $P_{\rm out}$ is computed from the output voltage $V_{\rm out} (t)$, which is reconstructed from the X and Y components. Accounting for known conversion factors and amplification settings of the rf lock-in amplifier.
As with dc sensor measurements, $P^{\rm rf}_{\rm diss}(t)$ yields an instantaneous value at each moment in time, which is then time averaged in order to find the entropy dissipated.

\section{\label{supp:tick_readout}Tick readout}
In this appendix, we detail the method following which the three states of the DQD were identified.
First, in Sec.~\ref{supp:identifying_states}, it is explained how the output signal from the sensor dot can be used to infer the DQD state.
Secondly, in Sec.~\ref{supp:empirical_precision}, we elaborate on how the precision $\mathcal S$ is found, and finally, in Sec.~\ref{supp:low_readout_regime}, we discuss the caveats of the regime where the readout signal to noise ratio (SNR) is too low to infer the DQD state.

\subsection{\label{supp:identifying_states}Identifying the three states}
Both measurement methods described in Sec.~\ref{supp:details_current} and~\ref{supp:details_reflectometry} yield a noisy signal as their output.
In case of the dc method, the output is a current expressed in units of $[I]={\rm pA}$.
For the rf reflectometry, the output is a voltage expressed in units of $[V]={\rm mV}$.
Figure~\ref{fig:endmatter} of the End Matter shows a comparison of two exemplary such traces.
For simplicity of presentation, we here refer to the current output signal $I_{\rm cs}(t)$ but the procedure for the analysis is the same for the rf sensor output signal. 
The key idea is to infer the charge state of the DQD from the instantaneous current value $I_{\rm cs}(t)$.
Generally, the current will be close to one of the three values $I_{\rm l}<I_{\rm m}<I_{\rm h}$ (see Fig.~\ref{fig:endmatter}), each corresponding to one of the three DQD states $\ket{0}$, $\ket{L}$ and $\ket{R}$.
Due to noise in the measured signal, the output current $I_{\rm cs}(t)$ will not necessarily match one of these values, therefore a real-time strategy has to be devised to determine which of the three values $I_{\rm l/m/h}$ (and thus which charge state of the DQD) is the most likely one.

The three current values can be determined by initially calibrating the setup.
In principle, the three current values could be determined theoretically, however, due to the non-trivial geometry of the device and the associated unknown parameters, the three values are determined empirically.
Here, we follow the thesis~\cite{Craig2023} for how these values are determined, and for the following description, we assume a fixed voltage bias $V_{\rm DQD}$ across the DQD and a fixed voltage bias $V_{\rm cs}$ across the sensor dot.
We furthermore assume that the SNR of the output signal is sufficiently high to clearly distinguish the three levels.
How to deal with the case of low SNR is discussed in Sec.~\ref{supp:low_readout_regime}.
For each experimental setting, a finite set of values $\mathcal I=\{I(t)_{\rm cs}\}_{t\in \mathcal T}$ is recorded within an approximately $30\,{\rm min}$ long time window $\mathcal T$.
The current values are then binned within a histogram giving a probability density function $p(I)$ that a certain value $I$ appears within the set $\mathcal I$ irrespective of time tag $t$ (see Fig.~\ref{fig:endmatter}(a,c) in the End Matter).
Since $I_{\rm cs}(t)$ generally takes one of the three values $I_{\rm l/m/h}$, the histogram will feature three peaks centered around those values.
We chose $m=180$ equidistant bins within the window $[I_{\min},I_{\max}]$ where $I_{\min} = \min \mathcal I$ and $I_{\max} = \max\mathcal I$, though for Fig.~\ref{fig:endmatter}, a lower number of bins is shown for readability.
The PDF $p(I)$ is then defined by
\begin{align}
    p(\tilde I_k) = \frac{|\{I_{\rm cs}(t)\in\mathcal I : I_k\leq I_{\rm cs}(t)< I_{k+1}\}|}{\Delta I\times |\mathcal I|},
\end{align}
where $I_k = k(I_{\max}-I_{\min}) /m + I_{\min}$ is the left boundary of the $k$th bin (and right boundary of the $(k-1)$st bin), ${\tilde I}_k \in [I_k,I_{k+1}]$ and $\Delta I = (I_{\max} - I_{\min})/m$.
For values outside the window $I\notin [I_{\min} , I_{\max }]$, the histogram PDF is set to zero $p(I)=0$.
As we see in Fig.~\ref{fig:endmatter}(a,c), the histogram shows three distinct peaks, both for the dc readout method and for the reflectometry method.

In a next step, we want to discretize the output signal $I_{\rm cs}(t)$ into a signal $S(t)$, where $S(t)$ can only take one of three values $0,R,L$ corresponding to the three DQD states.
To achieve this, we fit a the sum of three Gaussian distributions to the histogram data $p(I)$ with the intent that each of the three peaks from the histogram is well-matched by one of the Gaussians.
The parameters that have to be fit are given by the three means $\vec \mu$, widths $\vec \sigma$ and heights $\vec h$.
Note that the means $\vec \mu$ are what define the three calibrated values $I_{\rm l/m/h}$.
With all parameters, we can define the function
\begin{align}
    q_{\vec\mu,\vec\sigma,\vec h}(I) = \sum_{i\in\{0,R,L\}} \underbrace{\frac{h_i}{\sqrt{2\pi\sigma_i^2}} \exp\left(-\frac{\left(I-\mu_i\right)^2}{2\sigma_i^2}\right)}_{=:q_i(I)},
\end{align}
which is then fitted numerically with a least-squares algorithm (code available at~\cite{Git2025}).
We thus determine the three parameters according to
\begin{align}
\label{eq:fitted_gaussian_params}
    (\vec \mu^*,\vec \sigma^*,\vec h^*) = \underset{\vec \mu,\vec \sigma,\vec h}{\rm argmin} \sum_{k=1}^m |p(I_k) - q_{\vec \mu,\vec \sigma,\vec h}(I_k)|^2,
\end{align}
and identify $I_{\rm l}=\mu_{0}<I_{\rm m}=\mu_R<I_{\rm h}=\mu_L$ in the following ordering, using the charge stability diagram from Sec.~\ref{supp:details_current}.
This allows mapping the three current values to the DQD state as follows
\begin{align}
\label{eq:state_LMH_identification}
    \ket{0} \,\widehat{=}\,\,I_{\rm l}, \quad
    \ket{L} \,\widehat{=}\,\,I_{\rm h}, \quad
    \ket{R} \,\widehat{=}\,\,I_{\rm m}.
\end{align}
Up to small corrections in the normalization due to the finite window $[I_{\min},I_{\max}]$ and imperfect fit of $q_{\vec \mu,\vec \sigma,\vec h}$ to $p$, the three heights sum up to unity $h_1+h_2+h_3\approx 1$.
To formally ensure normalization, we can replace $h_i \mapsto h_i/(h_1+h_2+h_3)$.
Note that for the rf sensor output, the X and Y quadratures have analogous low/middle/high levels, however, for the PCA trace, middle and high levels are swapped.

\begin{figure}
    \centering
    \includegraphics[width=\linewidth]{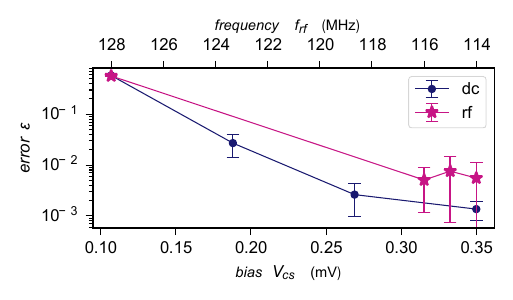}
    \caption{State identification error. As a function of sensor dot bias $V_{\rm cs}$ for the dc readout and frequency $f_{\rm rf}$ for the rf reflectometry readout, the average state identification error is shown.
    The average is taken over all DQD bias settings and the entire time-traces.
    The error bars show the sample standard deviation.
    Note that for the lowest readout fidelity setting (leftmost data points), the error shown is only an approximation of the true error due to the ambiguity of how the states are identified in this regime.}
    \label{fig:ID_error}
\end{figure}

Given this initial calibration, we intend to use it to classify the instantaneous current $I_{\rm cs}(t)$ with one of the three values $I_{\rm l/m/h}$ and hence conclude in which of the three states $\ket{0},\ket{R}$ or $\ket{L}$ the DQD was at time $t$.
The normalized functions $q_i(I)/h_i$ can be interpreted as the probability density that the current takes value $I$ conditioned on the system being in the state $i$.
To pick the most likely state for a given value of $I_{\rm cs}(t)$ we can thus simply take the value $i\in\{0,R,L\}$ for which $q_i(I)$ is maximal.
We thus define
\begin{align}
\label{eq:s_identification}
    S(t) = \underset{i\in\{0,R,L\}}{\rm argmax} \,q_i(I_{\rm cs}(t)),
\end{align}
and recalling the identification in eq.~\eqref{eq:state_LMH_identification}, we can interpret $S(t)$ as the instantaneous state of the DQD at time $t$.
The result of this identification is exemplarily shown in Fig.~\ref{fig:endmatter}(e) of the End Matter.
The overlap of the three Gaussians fitted to the histogram can be used to estimate the error of the state identification according to eq.~\eqref{eq:s_identification}.
If for a given output signal $I$, the state $S\in \{0,L,R\}$ is identified,
\begin{align}
\label{eq:varepsilon_S}
    \varepsilon(S) = 1-\frac{q_{S}(I)}{\sum_{i\in\{0,R,L\}}q_i(I)},
\end{align}
is an estimate for the error with which the state has been identified.
If some output value $I$ lies in the region where two Gaussians from the histogram fitting overlap, the probability $q_i(I)$ may be approximately be the same for those Gaussians, and hence, $\varepsilon \approx 1/2$, meaning there is an approximately 50\% chance that the state identification guess was wrong.
In Fig.~\ref{fig:ID_error} we plot the state identifiaction error for both dc and rf readout techniques, as average values over the entire time-traces and all the DQD voltage bias settings.
Note that in the low readout fidelity regime, the definition of the error in eq.~\eqref{eq:varepsilon_S} becomes ambiguous.
The value $\varepsilon(S)$ depends on the choice of how the Gaussians are fitted to the histogram which fails to be unique for the low fidelity regime.
In Sec.~\ref{supp:low_readout_regime}, this edge case is further discussed.

For a single measurement run and after initial calibration we thus obtain a time-series $S(t)$ for a discrete set of times $\mathcal T = [\tilde t_0,\tilde t_1,\dots,\tilde t_N]$.
The discrete times $\tilde t_i$ are given by the timing resolution of the measurement device, with $\tilde t_N-\tilde t_0 \approx 30\min$ and $N=111331$.
For simplicity we assume $\tilde t_0=0$.
In principle, one could of course use the clock from the readout device to obtain a much better timing estimate than from the DQD clock itself.
This is, however, only a technological caveat: for this work, the performance of the DQD is characterized against the (much better) clock in the readout device and therefore we know the the time tags $\tilde t_0,\tilde t_1,\dots,\tilde t_N$.
When using the DQD as a clock, it is not necessary to know these time tags; rather, one could simply convert the amplified analog reading of the instantaneous current $I_{\rm cs}(t)$ into the discrete state $S(t)$ and use the jumps between different states to obtain a time estimate.

\paragraph*{State identification ambiguity.}
When the current jumps from the high value $I_{\rm h}$ directly to the low value $I_{\rm l}$, it can generally be the case that the measurement device does not react quickly enough and thus records an intermediate value such that a naive post-processing would yield the sequence high-middle-low, instead of high-low.
To mitigate this problem, one can identify the middle current $I_{\rm m}$ with the state $\ket{R}$ only if the dc signal remains $k>3$ time steps in the middle region.
In our analysis all instances where $I_{\rm cs}(t)$ is close to $I_{\rm m}$ for only one or two data points, are rejected and instead identified as the previous state of $I_{\rm cs}(t)$.
However, when $k\geq 3$ neighboring values of $I_{\rm cs}(t)$ are close to $I_{\rm m}$, they are identified as a middle state.

\subsection{\label{supp:empirical_precision}Empirically estimating the precision}
Given the discretized time trace $S(t)$ it is now possible to calculate the time estimates $\Theta_{\rm net}$ and $\Theta_{\rm opt}$ proposed in the main text.
While $S(t)$ still encodes the state at all times within the recording window $\mathcal T$, an equivalent description of this time trace is the sequence of jumps $\mathbf s$ as introduced in the main text together with the jump times $\mathbf t$.
The sequences for the entire time trace are given by
\begin{align}
\label{eq:i_t_seq}
    \mathbf s = (s_0,s_1,s_2,\dots,s_m),\quad \mathbf t = (t_1,t_2,\dots,t_m),
\end{align}
with $s_k$ the state after the $k$th jump and $s_{k-1}$ the state before.
The time $t_k$ is the time of that $k$th jump.
Note, that there is no `$0$th' jump, and $m$ is the total number of jumps within the sequence.
These sequences can be obtained by taking the full time trace $(S(t_0),S(t_1),\dots,S(t_N))$ and identifying the indices $0\leq \ell_k\leq N-1$ where the two states $S(t_{\ell_k-1})\neq S(t_{\ell_k})$ differ.
We then take the state $S_k = S(t_{\ell_k})$ as the one after the jump and $t_k = \tilde t_{\ell_k}$ as the time of that jump.
With the sequence $\mathbf s$, the estimators $\Theta_{\rm net}$ and $\Theta_{\rm opt}$ can be constructed following the prescription in the main text.
To determine the estimators at time parameter $t$ (relative to the lab clock), one can consider the sub-sequence $\mathbf s(t) = \left(s_0,s_1,\dots,s_{m(t)}\right)\subseteq \mathbf s$ where $m(t)$ is the number of jumps at time $t$, characterized by $t_{m(t)}<t<t_{m(t)+1}$.
For convenience, we recall the definition of the estimators,
\begin{align}
    \Theta_{\rm net}[\mathbf s(t)] = \frac{1}{\nu}N[\mathbf s(t)],
\end{align}
where $N[\mathbf s(t)]$ is the net number of left-right charge transfers associated with the trajectory $\mathbf s(t)$, and
\begin{align}
\label{eq:Theta_opt_Supp}
    \Theta_{\rm opt}[\mathbf s(t)] = \sum_{k=0}^{m(t)} \frac{1}{\Gamma_{s_k}}.
\end{align}
We now proceed to detail how to explicitly construct the two estimators from the data.

\paragraph*{Construction of $\Theta_{\rm net}$.}
Let us start with how $\Theta_{\rm net}$ is determined.
We have identified $N[\mathbf s(t)]$ as the net number of charges that have tunneled across the DQD for a jump sequence $\mathbf s(t)$.
With the states $\ket{0}$ describing no charge occupation and $\ket{L/R}$ the left/right dot occupied respectively, a charge that has tunneled from the left to the right would correspond to the sequence
\begin{align}
\label{eq:LR_cycle}
    \ket{0}\rightarrow\ket{L}\rightarrow\ket{R}\rightarrow\ket{0}.
\end{align}
However, it may also be that the charge jumps back and forth between the left and right dot before it again leaves through the right dot, and thus a sequence
\begin{align}
    \ket{0}\rightarrow\ket{L}\rightarrow\ket{R}\rightarrow\ket{L}\rightarrow\ket{R}\rightarrow\ket{0},
\end{align}
would also constitute a valid left to right charge transport.
On the other hand, a charge transport from right to left is given by the inverse sequences,
\begin{align}
\label{eq:RL_cycle}
    \ket{0}\rightarrow\ket{R}\rightarrow\ket{L}\rightarrow\ket{0},
\end{align}
and so on.
Note that these cycles that transport a charge from the left to the right or vice-versa always have an even number of states, including the first and last state equals $\ket{0}$.
A cycle with with an odd number of transitions involves a charge entering and then leaving again through the same terminal: hence, this does not give rise to a net charge transfer.
The net number of charge transports $N[\mathbf s(t)]$ is thus defined as the number of left-to-right cycles (eq.~\eqref{eq:LR_cycle} and extensions) minus the number of right-to-left cycles (eq.~\eqref{eq:RL_cycle} and extensions).
We refer to the data availability statement regarding the precise algorithm used for implementing this counting.
See Fig.~\ref{fig:panel}(e) in the main text for a visualization for when the left-to-right versus right-to-left cycles occur in an examplary setting of the DQD.
The pre-factor $\nu$ is obtained by initially calibrating the DQD clock against a reference (e.g., lab time) to ensure the normalization of the time estimate, such that on average over all possible trajectories $\mathbf s(t)$, one has $\E[\Theta_{\rm net}[\mathbf s(t)]]=t+O(1)$ for large enough times $t$.

\paragraph*{Construction of $\Theta_{\rm opt}$.}
Next, we discuss the construction of the optimal estimator following the definition in eq.~\eqref{eq:Theta_opt_Supp}.
Here, the definition is more straightforward, as we can simply take eq.~\eqref{eq:Theta_opt_Supp} and rewrite
\begin{align}
\label{eq:Theta[i]_appendix}
    \Theta_{\rm opt}[\mathbf s(t)] = \sum_{s\in\{0,R,L\}} \frac{n_s(t)}{\Gamma_s},
\end{align}
where $n_s(t)=|\{s_k \in \mathbf s(t) : s_k=s\}|$ is the number of times the state $s$ appears in the sequence $\mathbf{s}(t)$.
Note that this includes both the initial state and the state the system occupies at time $t$, i.e., the state after the last jump.
The estimator as defined in eq.~\eqref{eq:Theta[i]_appendix} relies on knowledge of the jump rates $\Gamma_s$ that need to be estimated initially.
In the present experiment, as we show in Sec.~\ref{supp:characterizing_rates}, the rates are calibrated relative to a lab clock to express them in units of seconds.
However, as argued in~\cite{Prech2024}, the relative ratio of the rates may also be determined simply from the counting statistics without access to a lab clock.

\paragraph*{Precision estimate.}
With the prescriptions for how to obtain the two estimators, we are now in a position to estimate their precision $\mathcal S$ defined in eq.~\eqref{eq:S}.
To this end, we need to determine empirically the expectation value $\E[\Theta(t)]$ as well as $\Var[\Theta(t)]$.
With the sample sizes of $T=30\min$, one way to estimate the expectation value and the variance is to slice the time series $S(t)$ into $M$ sub-samples of duration $T/M$.
This defines $M$ new sequences
\begin{align}
    & \underbrace{s(0),s(\tilde t_1),\dots,s(\tilde t_{\lfloor{N/M}\rfloor})}_{=S_1},\\
    & \underbrace{s(\tilde t_{\lfloor{N/M}\rfloor+1}),\dots,s(\tilde t_{2\lfloor{N/M}\rfloor})}_{=S_2}, \\
    &\dots, \\
    &\underbrace{s(\tilde t_{(M-1)\lfloor{N/M}\rfloor+1}),\dots,s(\tilde t_{M\lfloor{N/M}\rfloor})}_{=S_M},
\end{align}
where we take each sub slice $S_k$ ($1\leq k\leq M$) to start at time $t=0$.
Following the prescription from eq.~\eqref{eq:i_t_seq}, the discretized jump sequences $\mathbf s, \mathbf t$ can be derived, giving us $M$ samples
\begin{align}
    S_k \mapsto (\mathbf s_k,\mathbf t_k).
\end{align}
Assuming the underlying DQD model follows the Markovian master equation~\eqref{eq:dotp=Mp}, these sequence are independent samples.
To good approximation, the experiment satisfies the Markovianity assumption in the regimes we probe; small deviations due to experimental imperfections are characterized in more detail in Sec.~\ref{supp:characterizing_rates}.
We thus treat these samples $(\mathbf s_k,\mathbf t_k)_k$ as independent and coming from an identical distribution.
These samples then allow us to empirically estimate the expectation value of the time estimator at time $t=T/M$ with
\begin{align}
\label{eq:ETheta_sample}
    \widehat\E [\Theta(t)] = \frac{1}{M}\sum_{k=1}^M\Theta[\mathbf s_k],
\end{align}
and the variance
\begin{align}
\label{eq:VarTheta_sample}
    \widehat\Var [\Theta(t)] = \frac{1}{M-1}\sum_{k=1}^M\big(\Theta[\mathbf s_k] - \langle\Theta(t)\rangle\big)^2.
\end{align}
We use the $\widehat{\,\cdot\,}$-notation to distinguish between the \textit{true} expectation value and variance and the \textit{sample} expectation value and variance given the finite sample size from the experiment.
Taking yet another average over all possible experimental runs should then again recover $\E[\widehat\E[\Theta(t)]] = \E[\Theta(t)]$, and similarly for the variance $\E[\widehat\Var[\Theta]]=\Var[\Theta]$.
The estimators are not of least variance but they are unbiased.
This general property does not require any assumption on the distribution of $\Theta(t)$ aside from the existence and finiteness of the first two moments of the distribution~\cite{Casella2024}.
The precision estimate we then define to be,
\begin{align}
\label{eq:S_hat}
    \widehat {\mathcal S}(t) = \frac{\widehat\E[\Theta(t)]^2}{t\, \widehat\Var[\Theta(t)]},
\end{align}
again recalling that $t=T/M$.
As a consequence, the larger the number of samples $M$, the shorter the time $t$.
Thus the sample precision $\widehat{\mathcal S}$ is further away from the asymptotic limit $t\rightarrow\infty$ of the definition of $\mathcal S$ in eq.~\eqref{eq:S}.
For the \textit{late-time} data shown in Fig.~\ref{fig:panel2} of the main text, we chose $M=300$ and thus $t\sim 7\sec$.

Note that the precision can also be used to infer the relative uncertainty of the normalization constant used to define the estimators.
Exemplarily, for the $\Theta_{\rm net}$-estimator, the frequency $\nu$ is initially calibrated relative to the lab clock.
The estimated frequency over a calibration time window $[0,t]$ is given by $\widehat{\nu} = N[\mathbf{s}(t)] / t$, and the relative uncertainty is approximated by
\begin{align}
    \frac{\widehat{\rm Var}[\widehat{\nu}]}{\widehat{\nu}^2} = \frac{1}{t\, \widehat{\mathcal S}},
\end{align}
and we emphasize that $\widehat{\mathcal S}$ does actually not depend on $\nu$.
Using the numerical values determined for the high DQD bias regime, with $\mathcal S\approx 1\,{\rm Hz}$ and $t\approx 30\,{\rm min}$, we find a relative uncertainty of around $2\%$.
We further remark that the uncertainty in $\mathcal S$ as analyzed in the following paragraph and shown in Fig.~\ref{fig:panel2} of the main text amounts to the uncertainty of the estimated calibration error $\widehat{\rm Var}[\widehat{\nu}]/\widehat{\nu}^2$.
As for the rates $\Gamma_s$ used to calibrate the optimal estimator $\Theta_{\rm opt}$, an error analysis can be found in Sec.~\ref{supp:characterizing_rates}.

\paragraph*{Error analysis.}
Given the sample precision $\widehat{\mathcal S}$, we need yet another estimate: the estimate for the error of the precision.
To obtain this, we first need to determine the standard error in the sample estimates $\widehat\E [\Theta(t)]$ and sample variance $\widehat\Var [\Theta(t)]$.
Assuming that for late times, $\Theta(t)$ becomes approximately normally distributed, the standard errors can be directly calculated following~\cite{Casella2024}.
Given $M$ iid\ samples $X_1,\dots,X_M$ from a Gaussian distribution with mean $\mu$ and variance $\sigma^2$, the sample mean and variance are defined akin to eqs.~\eqref{eq:ETheta_sample} and~\eqref{eq:VarTheta_sample} as $\widehat\E[X] = \frac{1}{M}\sum_{k=1}^MX_k$ and $\widehat\Var[X] = \frac{1}{M-1}\sum_{k=1}^M(X_k-\widehat\E[X])^2$.
For example in the textbook~\cite{Casella2024} it is then shown that the variance of these estimators are then given by $\Var[\widehat \E[X]]=\sigma^2/M$ for the sample mean, and the variance of the sample variance is given by $\Var[\widehat\Var[X]]=2\sigma^4/(M-1)$.
Note, that for the latter, the assumption that the $X_i$ are normally distributed is crucial; for more general distributions, the variance of the sample variance would depend on higher moments of the distribution.
Applying these results to our setting (recalling the assumption of $\Theta(t)$ being normally distributed), we do not know the underlying variance of $\sigma^2$ of the distribution.
Instead, we take the estimated variance $\widehat\Var[\Theta(t)]$ and thus find the estimated standard error of the sample mean,
\begin{align}
\label{eq:std_error_avg}
    \widehat\sigma_{\widehat\E [\Theta(t)]} = \frac{\sqrt{\widehat\Var [\Theta(t)]}}{\sqrt{M}}.
\end{align}
As for the variance, the estimated standard error is given by,
\begin{align}
\label{eq:std_error_var}
    \widehat\sigma_{\widehat\Var [\Theta(t)]} = \widehat\Var [\Theta(t)]\sqrt{\frac{2}{M-1}}.
\end{align}
To determine the standard error of the precision estimate from eq.~\eqref{eq:S_hat}, we use the Gauss error propagation formula which is valid for small standard-errors, and gives
\begin{align}
\label{eq:S_standard_error}
    \widehat\sigma_{\widehat {\mathcal S}(t)} 
    &= \frac{\widehat {\mathcal S}(t)}{\sqrt{M}}\sqrt{\frac{4\widehat\Var[\Theta(t)]}{\widehat\E[\Theta(t)]^2 } + \frac{2M}{M-1}}.
\end{align}
In Fig.~\ref{fig:panel2}, the error bars for the precision estimate display the error calculated according to eq.~\eqref{eq:S_standard_error}.
Finally, we note that this error analysis only includes the error coming from the inherent stochasticity of the estimator $\Theta(t)$ but not from erroneous identification of the states as outlined in Sec.~\ref{supp:identifying_states}.
Figure~\ref{fig:panel2}, however, shows that the precision estimate agrees within the error bars with the theoretical predictions given the rate matrix as we obtain in in the following Sec.~\ref{supp:characterizing_rates}, corroborating that the given error analysis includes the dominant sources of error.
Finally, we note that the measurement itself also has only negligible backaction on the clock, as can be seen in both Figs.~\ref{fig:panel2}(b,c), where the estimated precision varies little with measurement strength beyond the initial dissipation threshold, thus indicating that neither the measurement method nor the strength significantly affect the clock dynamics.

\subsection{\label{supp:low_readout_regime}Low readout fidelity regime}
For the present experiment, and every setting of the DQD voltage $V_{\rm DQD}$, four settings for the readout were used.
For the dc method, this was defined by the sensor bias $V_{\rm cs}$, and for the rf reflectometry method by the frequency of the input signal $f_{\rm rf}$ as detailed in Sec.~\ref{supp:details_reflectometry}.
Here, we discuss the analysis for the dc measurement, but the analysis generalizes without caveats to the rf reflectometry signal.

\begin{figure}
    \centering
    \includegraphics[width=\linewidth]{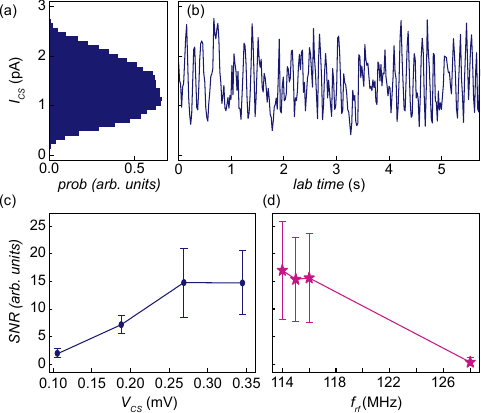}
    \caption{Example of a dc measurement trace for the lowest sensor dot bias setting together with a visualization of the SNR for the readout current.
    (a) Histogram for the current is shown for the settings $V_{\rm DQD}=0.455\,{\rm mV}$ and $V_{\rm cs}=0.102\,{\rm mV}$.
    We see that no three distinct levels are visible anymore in contrast to Fig.~\ref{fig:endmatter}(a,c).
    In (b), the current trace is shown for a time-window of approximately $5\,{\rm sec}$, further highlighting that no particular level structure emerges.
    (c, d) Readout SNR as calculated in eq.~\eqref{eq:SNR} is shown for the dc and rf sensing methods.
    Each data point is the average over all DQD biases $V_{\rm DQD}$ and the error bars indicate the sample deviation for the average over all DQD biases.}
    \label{fig:readout_SNR}
\end{figure}

Of the four sensor bias settings, three were in a regime where the three states were clearly distinguishable and thus it was possible to identify $I_{\rm cs}(t)\rightarrow S(t)$ according to eq.~\eqref{eq:s_identification}.
Working in the setting with lowest sensor bias, the state of the DQD can not be inferred from the current $I_{\rm cs}(t)$ anymore.
As we show in Fig.~\ref{fig:readout_SNR}(a,b), in the low sensor bias regime, the histogram of $I_{\rm cs}(t)$ resembles a normal distribution around some mean current value and thus the identification $S(t)$ according to the scheme outlined in Sec.~\ref{supp:identifying_states} fails.
There is no canonical choice for how to recover the three states from the noisy signal $I_{\rm cs}(t)$ in the lowest SNR setting.
One possibility is to identify the middle current $I_{\rm m}$ value with the mean of the histogram distribution from, e.g.\ Fig.~\ref{fig:readout_SNR}(a), and set $I_{\rm l}$ one standard deviation of the histogram below $I_{\rm m}$ and $I_{\rm h}$ one standard deviation above.
Note that the state identification based on this method does not necessarily allow determining the underlying state of the DQD.
In fact, due to the noise in the signal, it is not expected that the two coincide, which we can also see based on the following:
The net current estimator $\Theta_{\rm net}(t) \approx 0$ stays approximately at zero as Figs.~\ref{fig:panel2}(b,c) show.
The reason is that in the random pattern of $I_{\rm cs}(t)$ in the low sensor bias regime, there is no preferred cycle $0\rightarrow L\rightarrow R\rightarrow 0$ versus $0\rightarrow R\rightarrow L\rightarrow 0$ anymore in the identified signal $S(t)$, even though in the underlying DQD, the left-to-right cycles would be preferred.
The random occurrences of such cycles in both orientations cancel each other out and lead to no observed net count, even in the case where a biased charge transport is taking place in the underlying DQD.

Looking at the optimal estimator $\Theta_{\rm opt}(t)$ one may define every change in $I_{\rm cs}(t)$ above some small threshold as a tick of the clock, even though there is no underlying jump in the DQD.
The time we are reading in that case, however, is not the one given by the quantum clock, but the one given by time tags produced by the device recording the current.
It is thus no surprise that this way of defining the optimal estimator would result in a comparatively precise clock.
In the lowest sensor bias regime, we thus refrain from defining $\Theta_{\rm opt}(t)$ because $I_{\rm cs}(t)$ is not correlated with the DQD state anymore, and thus there is also no estimate for the number of jumps needed for $\Theta_{\rm opt}(t)$.

\paragraph*{Readout SNR.}
To more quantitatively assess the quality of the signal in the current trace $I_{\rm cs}(t)$ we can use the signal power to noise ratio.
In contrast to the state identification error as introduced in eq.~\eqref{eq:varepsilon_S}, the SNR does not heavily depend on the choice of how the states are identified from the signal.
Rather, it captures the fluctuations of the readout trace around the signal, and thus also yields an informative estimate of the readout quality in the low fidelity regime.
To this end, we may define the noise
\begin{align}
    \xi(t) = I_{\rm cs}(t) - I_S(t),
\end{align}
where $I_S(t)$ is the signal as identified with $S(t)$.
That is, $I_S$ takes one of the three values $I_{\rm l/m/h}$ depending on the respective state $S(t)=\{0,R,L\}$.
The noise power can then be calculated as the average squared noise,
\begin{align}
    \E[\xi^2] = \frac{1}{T}\int_0^T \dd t \xi(t)^2,
\end{align}
where $[0,T]$ is the recording time window.
In the case for the experiment, this would be approximated with a discrete Riemann sum from $\tilde t_0=0$ to $\tilde t_N\approx 30\min$.
The signal power in turn is given by
\begin{align}
    \Var[I_s] = \frac{1}{T}\int_0^T \dd t \left(I_S(t) - \E[I_S]\right)^2,
\end{align}
with $\E[I_S]=1/T \int_0^T\dd t I_S(t)$.
We thus define the SNR of the readout current $I_{\rm cs}(t)$ as
\begin{align}
\label{eq:SNR}
    {\rm SNR} := \frac{\Var[I_S]}{\E[\xi^2]}.
\end{align}
In Fig.~\ref{fig:readout_SNR}(c), we show how the readout SNR behaves as a function of the sensor dot bias, revealing that for the lowest voltage setting, the SNR goes to approximately 3, whereas for the higher voltage settings the SNR is between 10 and 20.
The reason, the SNR as defined in eq.~\eqref{eq:SNR} only goes down to a value of 3 is that $I_S$ distinguishes three levels, so the noise $\xi$ is at most 1/3 of the fluctuation of the full current $I_S$.
Similarly, Fig.~\ref{fig:readout_SNR}(d) shows the readout SNR as a function of the frequency $f_{\rm rf}$ of the rf input signal for the reflectometry method.

\begin{figure*}
    \centering
    \includegraphics[width=\linewidth]{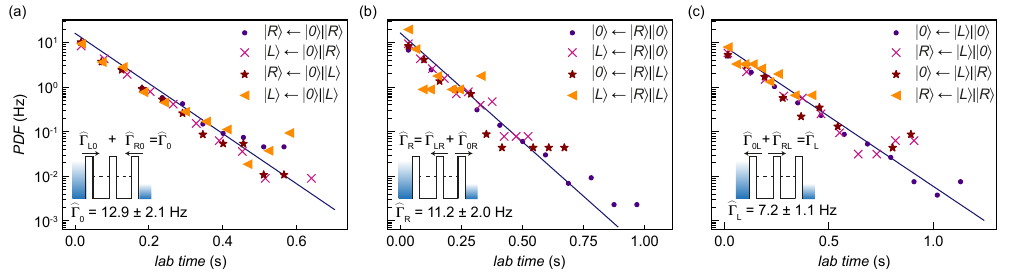}
    \caption{An representative visualization of the jump waiting times, here for the setting $V_{\rm DQD}=0.455\,{\rm mV}$ and $V_{\rm cs}=0.345\,{\rm mV}$.
    Each of the three subplots (a), (b) and (c) shows the waiting time for leaving one of the three states $\ket{0}$, $\ket{R}$ or $\ket{L}$ respectively.
    Four dot markers with identifier $j\leftarrow i|k$ show the histogram density value for the time the system stays in state $i$ before jumping to state $j$ and after having been in $k$ before.
    The histogram is taken with $n=10$ equidistant bins between the minimum and maximum value of the observed jump times.
    The number of observed waiting times ranges between $\sim 40$ and $\sim 4600$.
    Due to the Markovian dynamics of the master equation, these waiting times should all be distributed according to the same waiting time, as shown by the solid line.
    The solid line shows the exponential curve $\Gamma \exp(-\Gamma(t-\delta))$ accounting for the deadtime $\delta$, which is in agreement with the histogram data points.
    Some of the deviations between the data points and the solid line are due to the low numbers of observed sequences, e.g., the sequence $L\leftarrow R | L$ occurred approximately only once per minute giving 38 waiting times which is not significantly larger than the number of bins $n=10$.}
    \label{fig:cond_rates}
\end{figure*}
\section{\label{supp:characterizing_rates}Fitting the rate matrix}
In this Appendix, we discuss how the rate matrix $M$ describing the stochastic evolution of the system as in eq.~\eqref{eq:dotp=Mp} can be obtained from the measurement statistics. 
In Sec.~\ref{supp:maximum_likelihood_estimation} we derive the maximum likelihood estimate for the jump rates within the system, following which we go into a more detailed error analysis in Sec.~\ref{supp:extended_error_analysis}.
Then, Sec.~\ref{supp:recovering_full_rate_matrix} details how the rate matrix $M$ can be obtained as a global maximum likelihood estimate, showing consistency with the method discussed in the previous two sections.
Section~\ref{supp:theoretical_precision} details how the theoretical prediction of the precision $\mathcal S$ is obtained from the rate matrix $M$.
Finally, Sec.~\ref{supp:rate_drifts} discusses how drifts in the rates may lead to deviations between the theoretical predictions of the master equation model and experimental observations.
For the following discussions, we assume to work with a fixed setting of $V_{\rm cs}$ and $V_{\rm DQD}$ for the experiment, again in the regime with readout SNR high enough such that the current $I_{\rm cs}(t)$ can be discretized to the signal $S(t)$ with sufficiently small error.

\subsection{\label{supp:maximum_likelihood_estimation}Jump rate maximum likelihood estimation}
Given the discretized signal and following the procedure in Sec.~\ref{supp:empirical_precision}, we can obtain the sequences $(\mathbf s,\mathbf t)$ like in eq.~\eqref{eq:i_t_seq}.
Note that the entries $s_k$ denote the state after the $k$th jump and $t_k$ is the time of the $k$th jump.
Correspondingly, $s_{k-1}$ is the state just before the $k$th jump.
From this sequence, it is possible to extract the statistics on the time the system rests in, for example, state $i$ before it jumps to state $j\neq i$.
Let us denote with $T_{j|i}$ the random variable describing the time the system remains in $i$ before it jumps to $j$.
The sequences $(\mathbf s,\mathbf t)$ allow us to obtain a list of times,
\begin{align}
\label{eq:tau_tag_list}
    \mathcal T_{j|i} = \left(\tau_{j|i}^1,\tau_{j|i}^2,\dots,\tau_{j|i}^{N_{j|i}}\right),
\end{align}
where each entry is given by the difference $\tau_{j|i}^k=t_{\ell_k}-t_{\ell_k-1}$ of two neighboring entries in $\mathbf t$, where $i_{\ell_k}=j$ is the state after the jump and $i_{\ell_k-1}=i$ the state before the jump.
The indices $\ell_k$ are some dummy-label to index those states within the sequences $(\mathbf s,\mathbf t)$ that correspond to such a jump from $i$ to $j$.

\paragraph*{Waiting time distribution.}
Let us consider the classical master equation model~\eqref{eq:dotp=Mp} with rate matrix $M$.
The off-diagonal entries $\Gamma_{ji}:=M_{ji}$ are the rates with which the system transitions from state $i$ to $j$ and the diagonal entries $\Gamma_i:=-M_{ii}$ describe the waiting times.
The jump times $T_{j|i}$ should then be exponentially distributed, following
\begin{align}
\label{eq:P[T_ji]}
    P[T_{j|i}=\tau] = \Gamma_{i}e^{-\Gamma_{i}\tau}.
\end{align}
This can be shown by explicitly writing
\begin{align}
\label{eq:P[T_ji]_derivation}
    P[T_{j|i}=\tau] &= W[\tau, j | j\leftarrow i] \\
    &= \frac{W[\tau, j | i]}{P[j | i]},
\end{align}
and $W[\tau, j | i]=\Gamma_{ji}e^{-\Gamma_i \tau}$ is the probability density that the system jumps from $i$ to $j$ at time $\tau$ as calculated with the master equation $\dot{\mathbf p}=M\mathbf p$.
Furthermore, $P[j|i]=\Gamma_{ji}/\Gamma_i$ is the probability that the system jumps to $j$ after starting in $i$ (irrespective of the time), which also follows from the master equation.
What we can already discern at this stage is that the waiting time distribution for $T_{j|i}$ is independent of the state $j$ to which the system jumps.
This is a general property of classical master equations, and as we show in Fig.~\ref{fig:cond_rates}, the distribution of the jump times $\mathcal T_{j|i}$ does not depend on the state $j$.
A more quantitative assessment of this can be obtained by estimating the jump rate $\Gamma_i$ from the data set $\mathcal T_{j|i}$ and checking whether it differs between different choices of $j$.

\paragraph*{Maximum likelihood estimator for $\Gamma_i$.}
Assuming that the times $\mathcal T_{j|i}$ are indeed distributed according to eq.~\eqref{eq:P[T_ji]}, but $\Gamma_i$ is unknown, we now attempt to determine $\Gamma_i$ based on the data set $\mathcal T_{j|i}$.
One method of estimating $\Gamma_i$ is to use the maximum likelihood estimator (MLE) method~\cite{Casella2024}.
There, the estimator $\hat\Gamma_i$ is defined as the parameter that maximizes the likelihood of observing the given jump times $\mathcal T_{j|i}$,
\begin{align}\label{eq:Gamma_hat}
    \widehat \Gamma_i := \underset{\Gamma_i}{\mathrm{argmax}}\, P[\tau_{j|i}^1,\dots,\tau_{j|i}^{N_{j|i}}].
\end{align}
It is important to note, that in general, the MLE is not unbiased for finite samples, i.e., $\E[\widehat\Gamma_i]\neq \Gamma_i$ for $N_{j|i}<\infty$, and it may also not minimize the variance $\Var[\widehat\Gamma_i]$.
Yet, it is the most likely rate to reproduce the given dataset, and keeping these caveats about the MLE in mind, we now proceed to explicitly determine $\widehat\Gamma_i$.
For the present model, $P[\tau_{j|i}^1,\dots,\tau_{j|i}^{N_{j|i}}]=\prod_{k=1}^{N_{j|i}} \Gamma_i e^{-\Gamma_i \tau_{j|i}^k}$ is the likelihood of having observed the given dataset.
Here it is advantageous to consider the log-likelihood function
\begin{align}
\label{eq:log_likelyhood}
    \log P[\tau_{j|i}^1,\dots,\tau_{j|i}^{N_{j|i}}] = N\log \Gamma_i - \Gamma_i\sum_{k=1}^{N_{j|i}}\tau_{j|i}^k.
\end{align}
Since the log-function is strictly monotonously growing, the argument maximizing the log-likelihood function~\eqref{eq:log_likelyhood} equals the one maximizing of the likelihood function as in eq.~\eqref{eq:Gamma_hat}.
The MLE can thus be directly calculated to be
\begin{align}
    \widehat\Gamma_i = \frac{N_{j|i}}{\sum_{k=1}^{N_{j|i}} \tau_{j|i}^k}.
\end{align}

\paragraph*{Preliminary error analysis.}
If we assume for the moment that the times $\mathcal T_{j|i}$ are indeed distributed according to eq.~\eqref{eq:P[T_ji]} we can calculate the expectation value and the variance of the estimator $\widehat \Gamma_i$ analytically.
Ultimately, we want to use this to determine how closely $\widehat \Gamma_i$ approximates the true value of $\Gamma_i$.
Starting with the expectation value, it can be directly calculated using the likelihood function $P[\tau_{j|i}^1,\dots,\tau_{j|i}^{N_{j|i}}]$, which gives,
\begin{align}
\label{eq:EGamma}
    \mathrm{E}[\widehat\Gamma] &= \int_{\vec \tau\geq 0} \dd \vec \tau \left(\prod_{i=1}^N \Gamma e^{-\Gamma \tau^i}\right)\frac{N}{\sum_{i=1}^N \tau^i}.
\end{align}
For readability, the subscripts from the jump times $\tau_{j|i}^k$ have been omitted and we use $\vec \tau = (\tau^1,\dots,\tau^N)$ as a shorthand.
Moreover, $N$ is the number of jumps which was previously labelled as $N_{j|i}$ and we simply write $\Gamma$ instead of $\Gamma_i$ because the calculation is the same for all states $i=1,2,3$.
Since the integrand is of the form $f(\tau^1 + \cdots + \tau^N)$, the $N$-dimensional integral can be simplified to be a 1-dimensional one,
\begin{align}
\label{eq:integral_symmetry}
    \int_{\vec \tau\geq 0} \dd \vec \tau f\left(\sum_{i=1}^N \tau^i\right) &= \frac{N^{N/2}}{(N-1)!}\int_0^\infty \dd r\, r^{N-1} f\left(\sqrt{N}r\right).
\end{align}
This simplification can be realized using the $N$-dimensional substitution rule and introducing the new coordinate $r=(\tau^1 +\cdots + \tau^N)/\sqrt{N}$.
Then, we can write
\begin{align}
    \int_{\vec \tau\geq 0} \dd \vec \tau f\left(\sum_{i=1}^N \tau^i\right) &= V_N \int_0^\infty dr \left(r\sqrt{N}\right)^{N-1} f\left(\sqrt{N}r\right),
\end{align}
where $V_N$ is the area of the $(N-1)$-dimensional standard simplex embedded in $N$-dimension with unit distance from the origin.
The area can be calculated analytically, to be given by
\begin{align}
    V_N &= \mathrm{Vol}_{N-1}\left\{ \begin{bmatrix}
\tau^1 \\
\vdots \\
\tau^N\end{bmatrix} \Big| \sum_{i=1}^N \tau^i = 1, \tau^i \geq 0\right\} \\
&=\frac{\sqrt{N}}{(N-1)!},
\end{align}
as for example in Ref.~\cite{Stein1966}.
Using the simplified form for the integral~\eqref{eq:EGamma}, we can explicitly calculate
\begin{align}
\label{eq:E_HatGamma}
    \mathrm{E}[\widehat\Gamma] &= \frac{N^{N/2}}{(N-1)!}\sqrt{N}\int_0^\infty \dd r\, r^{N-2}e^{-\Gamma\sqrt{N}r} \\
    &= \frac{N}{N-1}\Gamma, 
\end{align}
for the first moment.
This shows that the MLE for the jump rate $\Gamma$ is biased, though asymptotically, the bias vanishes as $O(N^{-1})$.
As for the second moment, we can calculate
\begin{align}
    \mathrm{E}[\widehat\Gamma^2] &= \frac{N^{N/2}}{(N-1)!}N \int_0^\infty dr \Gamma^N r^{N-3}e^{-\Gamma\sqrt{N}r} \\
    &= \frac{N^2}{(N-1)(N-2)}\Gamma^2
\end{align}
Thus, the variance of the estimator is given by
\begin{align}
    \mathrm{Var}[\widehat\Gamma] &= \frac{1}{(N-2)} \frac{N^2}{(N-1)^2}\Gamma^2 \\
    &= \frac{1}{N}\Gamma^2 + O(N^{-2}).
    \label{eq:Var_HatGamma}
\end{align}
Based on this analysis we would thus expect the variance of the estimator $\widehat\Gamma$ to decrease with $\Gamma^2/N$ to leading order as $N\rightarrow\infty$.

\begin{figure}
    \centering
    \includegraphics[width=\linewidth]{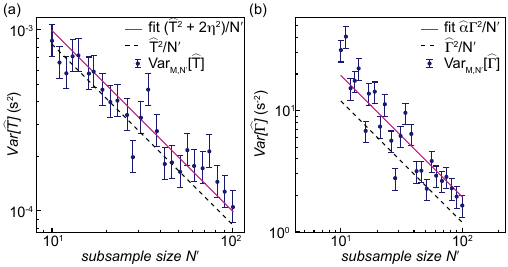}
    \caption{Showcase of the bootstrapping error analysis for the waiting times, with data from the settings $V_{\rm DQD}=0.455\,{\rm mV}$ and $V_{\rm cs}=0.345\,{\rm mV}$.
    The two plots are based on the times $\mathcal T=(t^1,\dots,t^N)$ the system has stayed in state $\ket{0}$ before jumping to another state.
    For the bootstrapping analysis $M=50$ pseudorandom subsets $\mathcal T^1,\dots,\mathcal T^M\subseteq \mathcal T$ each of length $N'$ are selected.
    (a) Here, we show the sample variance taken over the $M$ samples, for the survival time estimator $\widehat T$ as defined in eq.~\eqref{eq:HatT} as a function of the sample size $10\leq N'\leq 100$.
    The dashed line shows the theoretical scaling $\widehat T^2 /N'$ if there is no intrinsic uncertainty in the survival time, and the solid line shows the fit in case an intrinsic uncertainty $\eta$ of the survival is assumed as in the model eq.~\eqref{eq:PT_full}.
    (b) Sample variance is shown for the maximum likelyhood estimator $\widehat\Gamma$ as defined in eq.~\eqref{eq:HatVar_MN_HatGamma}.
    It can also be seen here that the fitted sample variance (solid line) decays less quickly than the theoretical scaling $\widehat\Gamma^2/N'$ shown with the dashed line.
    The reason is that, aside from the inherent stochasticity of the decay process, the rate is also not perfectly stable (see Sec.~\ref{supp:rate_drifts}).
    In both panels, the error bar indicates the estimated standard error of the sample variance, assuming Gaussian distributed errors, which underestimates the true error including further sources and systematic effects from the bootstrapping method.}
    \label{fig:bootstrap}
\end{figure}

To test whether this idealized model for the jump times describes well the experimentally observed jump statistics, we can check, whether the estimator indeed obeys the scaling from eq.~\eqref{eq:Var_HatGamma}.
To check this, we not only need to calculate the estimator $\widehat\Gamma$ but we also need to estimate the variance of the estimator.
At the same time, we are restricted by the data set at hand which comprises a finite set of experimentally observed jump times $\mathcal T = (\tau^1,\dots,\tau^N)$.
One way of estimating the variance of the estimator as a function of the sample size $N$ is the method of bootstrapping~\cite{Casella2024}.
For this, we take $M$ randomly selected subsets $\mathcal T^1,\dots\mathcal T^M\subseteq \mathcal T$ each comprising $N'\ll N$ jump times from $\mathcal T$ without any duplicates.
For each of the $M$ subsets we then determine the estimator
\begin{align}
\label{eq:HatGamma_Tk}
    \widehat\Gamma_{\mathcal T^k} = \frac{N'}{\sum_{\tau\in\mathcal T^k}\tau}.
\end{align}
This defines a sample of $M$ estimators $(\widehat\Gamma_{\mathcal T^1},\dots,\widehat\Gamma_{\mathcal T^M})$ for which we can calculate both sample average of the estimator
\begin{align}
\label{eq:HatE_MN_HatGamma}
    \widehat\E_{M,N'}[\widehat\Gamma]=\frac{1}{M}\sum_{k=1}^M \widehat\Gamma_{\mathcal T^k}
\end{align}
and the sample variance,
\begin{align}
\label{eq:HatVar_MN_HatGamma}
    \widehat\Var_{M,N'}[\widehat\Gamma]=\frac{1}{M-1}\sum_{k=1}^M \left(\widehat\Gamma_{\mathcal T^k} - \widehat\E_{M,N'}[\widehat\Gamma]\right)^2.
\end{align}
Under the assumption that the $M$ sets $\mathcal T^1,\dots,\mathcal T^M$ are independent and identically sampled sets of jump times according to the distribution~\eqref{eq:P[T_ji]}, we have the following.
The true expectation value of the sample average from eq.~\eqref{eq:HatE_MN_HatGamma} equals~\eqref{eq:E_HatGamma}, and the true expectation value of the sample variance from eq.~\eqref{eq:HatVar_MN_HatGamma} equals~\eqref{eq:Var_HatGamma} (with $N$ replaced by $N'$).
Due to finiteness of the data set $\mathcal T$ and finite sample size $M$ the subsets $\mathcal T^1,\dots,\mathcal T^M$ are of course not independent, and due to experimental deviations from the master equation model~\eqref{eq:dotp=Mp}, the samples will also not be sampled perfectly identically from the distribution in eq.~\eqref{eq:P[T_ji]}.
Keeping this caveat in mind, we visualize in Fig.~\ref{fig:bootstrap} how the sample estimate and variance from eqs.~\eqref{eq:HatE_MN_HatGamma} and~\eqref{eq:HatVar_MN_HatGamma} scale as a function of $N'$.
While the numerical analysis shows agreement of the sample average with the estimate from the full data set, the sample variance~\eqref{eq:HatVar_MN_HatGamma} shows a constant factor offset from this predicted behavior of eq.~\eqref{eq:Var_HatGamma}, meaning it scales inverse linearly with $N'$ but the slope is not $\widehat \Gamma^2$ but $\alpha\widehat \Gamma^2$ with $\alpha>1$.
In the following, we investigate the following two explanations for why the sample variance systematically exceeds the predicted estimator variance:
\begin{itemize}
    \item Firstly, detector deadtime will lead to deviations from the exponential distribution from eq.~\eqref{eq:P[T_ji]}.
    Since the measurement traces are taken at some finite equidistant time intervals the jump events that occur faster than the deadtime are not captured by this method of readout.
    \item Secondly, we do not expect the jump rates $\Gamma_i$ to be stable over the course of the experimental run.
    Thus, a more appropriate model assigns intrinsic uncertainty $\delta\Gamma_i$ to the underlying jump rate $\Gamma_i$.
\end{itemize}
To accurately estimate the variance of the rate estimator $\widehat\Gamma$ we thus resort to numerically fitting $\alpha$ to the scaling
\begin{align}
\label{eq:HatVar_MN_HatGamma_approxAlpha}
    \widehat\Var_{M,N'}[\widehat\Gamma] \sim \frac{\alpha \widehat\Gamma^2}{N'},
\end{align}
for $N'$ within the range $10 \leq N' \leq 100$ and $M=50$, as shown in Fig.~\ref{fig:bootstrap}(b).
Then we extrapolate the expression to $N>N'$ to obtain the variance of the estimator.
For estimating the error of the $\widehat\Gamma$, however, we also need to account for the intrinsic uncertainty $\hat\delta\Gamma$ in the rate, which does not vanish as $N$ grows.
The error estimate is thus given by
\begin{align}
\label{eq:HatSigma_HatGamma}
    \widehat\sigma_{\widehat\Gamma} := \sqrt{\frac{\alpha \widehat\Gamma^2}{N} + \hat\delta\Gamma^2},
\end{align}
where the two variances are summed together.
The additional term $\hat\delta\Gamma$ is the estimated intrinsic error of the decay rate, that we obtain in the following Sec.~\ref{supp:extended_error_analysis}.

\subsection{\label{supp:extended_error_analysis}Extended error analysis rate estimator}
In this section, we discuss how the estimator $\widehat\Gamma$ can take into account possible deadtime in the readout and furthermore, we analyze intrinsic errors in the decay rates.

\paragraph*{Detector deadtime.}
Even though the underlying Markov process for the evolution of the DQD's state follows the distribution in eq.~\eqref{eq:P[T_ji]}, there may be some jumps that occur to fast to be detected by the present measurement.
The reason for this can be that the state changes so quickly that the current signal $I_{\rm cs}(t)$ does not react at all or in other instances, it may be that the current signal responds to the state change but the identification $I_{\rm cs}(t) \mapsto S(t)$ does not capture this because the jump was too quick.
We can exemplarily look at a transition between states $i\rightarrow j\rightarrow k$ with $i,j,k$ all distinct with jump times $t_1<t_2$.
When the transition $j\rightarrow k$ happens too quickly for the sensor to register the jump, only the sequence $i\rightarrow k$ with a single jump time $t\equiv t_1\approx t_2$ will be recorded.
As a consequence:
\begin{itemize}
    \item The survival time $j\rightarrow k$ is overestimated because the fast events are not registered.
    \item Since the survival time in state $i$ is independent of the target of the jump, the survival time statistics for state $i$ will not be affected because $t_1\approx t_2$.
    \item However, the relative rates at which the system jumps $i\rightarrow k$ versus $i\rightarrow j$ will be affected, because the jump $i\rightarrow j$ is wrongly counted as a jump $i\rightarrow k$.
    If a transition $j\rightarrow k$ is thus much faster than the deadtime, this will lead to overcounting transitions $i\rightarrow k$ for other states $i\neq j$.
\end{itemize}
For our purposes, we restrict to the first effect, namely the shift in the survival time distribution due to the correction with the deadtime.
To address this, we work with a shifted exponential distribution, and for simplicty we again drop the state label.
Take $T$ to be the time between jumps, and $\Gamma$ the rate of the decay process.
Furthermore, take $\Delta$ to be the detector deadtime, then
\begin{align}
    P[T=\tau] = \Gamma \Theta[\tau \geq \delta] e^{-\Gamma (t-\delta)},
\end{align}
is the probability that a jump event is observed after time $\tau$.
The MLE in this case can be found by maximizing the log-likelihood function as in Sec.~\ref{supp:maximum_likelihood_estimation}, with a correction due to $\delta$.
If we also take $\delta$ to be an unknown, we can find the joint MLE for $\Gamma$ and $\delta$ with the solution that
\begin{align}
    \widehat\delta = \min_{\tau\in\mathcal T}\tau,
\end{align}
the estimated dead time is the smallest observed jump time.
The estimated jump rate is then given by,
\begin{align}
\label{eq:hatGamma_full}
    \widehat\Gamma = N/\sum_{\tau\in\mathcal T}\left(\tau -\widehat\delta\right).
\end{align}
As before $\mathcal T=(\tau^1,\dots,\tau^N)$ denotes the set of jump times obtained through the experiment, and the subscript denoting which transition they correspond to are omitted.

\paragraph*{Intrinsic rate error.}
As for the randomness of the jump rates, we can incorporate this by sampling the jump rate from a distribution for each jump.
For reasons of mathematical convenience, we switch to the waiting time picture where instead of sampling randomly the jump rate $\Gamma$ we randomly sample the survival time $\Delta=1/\Gamma$.
Furthermore, we chose a heuristic model where $\Delta$ is sampled from a normal distribution with average $\Delta_0$ and variance $\eta^2$,
\begin{align}
\label{eq:P[Delta]}
    P[\Delta] &= \mathcal N(\Delta_0,\eta^2)[\Delta] \\
    &= \frac{1}{\sqrt{2\pi \eta^2}}\exp\left(-\frac{(\Delta-\Delta_0)^2}{2\eta^2}\right).
\end{align}
Our goal is to determine the impact of this intrinsic uncertainty in $\Delta$ on the survival time statistics from the experimental data $\mathcal T$.
We first characterize the effects of this additional source of uncertainty in the theoretical model and then we provide a comparison with the experimental data to see how well the present model captures the actual data.

The probability distribution for the jump can now be calculated by the marginalizing the joint probability as follows,
\begin{align}
\label{eq:PT_full}
    P[T=\tau] &= \int \dd \Delta P[\Delta] P[T=\tau |\Delta].
\end{align}
The conditional probability distribution $P[T=\tau|\Delta]$ still includes the dead-time $\delta$ and is given by
\begin{align}
    P[T=\tau|\Delta] = \Theta[\tau\geq\delta]\frac{e^{-(\tau-\delta)/\Delta}}{\Delta}.
\end{align}
For simplicity, we assume that for each jump, the instantaneous survival time $\Delta$ is independently and identically sampled from the normal distribution $P[\Delta]$.
This assumption captures white noise in the survival time.
It is important to keep in mind that drifts in the experimental parameters may lead to correlated errors, where the survival times $\Delta$ for jump events close to each other may be closer than the survival time for jump events that happen far apart.
Such drifts in the survival time are investigated further in Sec.~\ref{supp:rate_drifts}.
To estimate the time between jumps $T$, we can consider
\begin{align}
\label{eq:HatT}
    \widehat T = \frac{1}{N} \sum_{\tau \in \mathcal T} \tau
\end{align}
as our estimator.
Note that $\widehat T$ is not necessarily an estimator for $\Delta$ (or $\Delta_0$) due to the dead time $\delta$.
Under the assumption that $\tau\in\mathcal T$ are independently and identically distributed jump times according to eq.~\eqref{eq:PT_full}, we can directly calculate
\begin{align}
    \mathrm{E}[\widehat T] &= \frac{1}{N}\sum_{i=1}^N\int \dd \tau \dd \Delta P[\Delta]P[T=t|\Delta]\tau \\
    &= \int \dd \Delta P[\Delta] \left(\Delta + \delta\right)\\
    &= \Delta_0 + \delta.
\end{align}
As for the second moment, we first notice that $\widehat T^2$ has $N$ terms of the form $\tau_i^2$ and $N^2-N$ terms of the form $\tau_i\tau_j$ with $i\neq j$.
Thus, $\E[\widehat T^2]$ splits into $N$ times the expectation value $\E[\tau^2]$ and $N^2-N$ times the expectation value squared $\E[\widehat T]^2$, both normalized by the prefactor $N^{-2}$.
In summary, we find,
\begin{align}
    \mathrm{E}[\widehat T^2] &= \frac{1}{N}\int \dd \tau \dd \Delta P[\Delta]P[T=t|\Delta] \tau^2 + \frac{N-1}{N}\mathrm{E}[\widehat T]^2 \nonumber \\
    &= \frac{1}{N}\int \dd \Delta P[\Delta]\left(2\Delta^2 + 2\Delta\delta + \delta^2\right) + \frac{N-1}{N}\E[\widehat T]^2 \nonumber \\
    &= \frac{1}{N}\left(\Delta_0^2 + 2\eta^2\right) + (\Delta_0+\delta)^2.
\end{align}
In combination together with the first moment, we can therefore determine the variance of the waiting time estimator $\widehat T$ to be given by
\begin{align}
\label{eq:Var_HatT}
    \mathrm{Var}[\widehat T] = \frac{\Delta_0^2 + 2\eta^2}{N}.
\end{align}
In the variance for $\widehat T$, the dead time $\delta$ drops out, and what remains are the two contributions, where one comes from the intrinsic survival time uncertainty $\eta$, and the other from the stochasticity of the jump events, adding the uncertainty $\Delta_0$.
With a bootstrapping method like how we analyzed the jump rate statistics in eqs.~(\ref{eq:HatGamma_Tk}--\ref{eq:HatVar_MN_HatGamma}), it is also possible to numerically verify the scaling of $\mathrm{Var}[\widehat T]$ with $N$ as in eq.~\eqref{eq:Var_HatT}.
Indeed, as we show in Fig.~\ref{fig:bootstrap}(a) for a representative data set for the jump times $\mathcal T$, we find that there is a correction $\eta>0$ needed to explain the variance of $\widehat T$.
This corroborates how the stochastic jump rates are not perfectly stable over the course of the experiment and subject to slight perturbations.
Furthermore, this factors into the estimated error for the rate estimator $\widehat\Gamma$.
The estimated error of the rate estimator comprises of two contributions, one coming from the inherent stochasticity of the jump process and thereby the statistical distribution of the waiting times.
This first contribution, however, shrinks with the number of samples $N$.
The second error we take into account is the intrinsic uncertainty in the rate which we can estimate with $\eta$.
The uncertainty in the survival time $\Delta$ is given by a Gaussian distribution with width $\eta$ according to eq.~\eqref{eq:P[Delta]}.
Decay rate and survival time are related by $\Gamma=1/\Delta$, and thus, if $\eta \ll \Delta$, we can approximate the uncertainty of the waiting time by $\Gamma (\eta / \Delta_0)$.
It turns out that $\eta/\Delta_0$ is approximately equal $10\%$ to $30\%$ and thus we can estimate the intrinsic rate uncertainty as
\begin{align}
    \hat\delta\Gamma = \widehat\Gamma^2 {\eta},
\end{align}
giving the expression to fit into the estimated rate error~\eqref{eq:HatSigma_HatGamma}.

\begin{figure*}
    \centering
    \includegraphics[width=\textwidth]{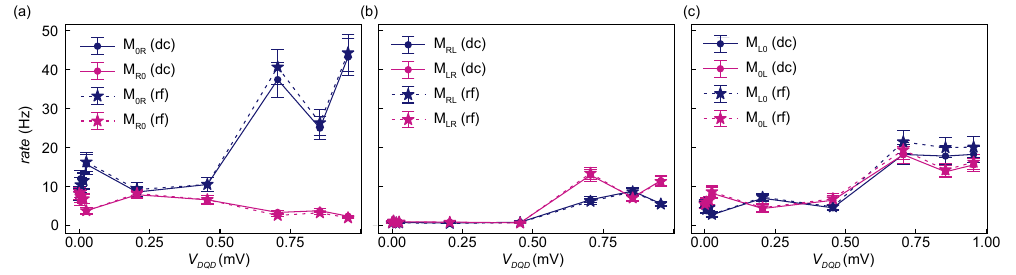}
    \caption{Estimated matrix elements $M_{ij}$ for $j\neq i$ describing the double quantum dot in the various voltage bias regimes.
    The $y$-axis shows the value of the matrix elements in units of ${\rm Hz}$ and the $x$-axis shows the DQD bias $V_{\rm DQD}$ in units of mV.
    We are showing the data obtained from both dc and rf reflectometry measurement from the same experimental run.
    The sensor bias was set to $V_{\rm cs} = 0.345\, {\rm mV}$ for the dc measurement and $f_{\rm rf}=114\,{\rm MHz}$ for the rf method, and thus the readout SNR is maximal in both cases.
    Over the three panels we note that around $V_{\rm DQD}=$ $0.5 \,{\rm mV}$, there is a large change in the magnitude of the tunneling rates reflecting the charge transport threshold.
    The error bars indicate the error in the estimated rates as estimated by the expression in eq.~\eqref{eq:HatSigma_HatGamma}.}
    \label{fig:rates_vs_DQD_voltage}
\end{figure*}

\subsection{\label{supp:recovering_full_rate_matrix}Recovering the full rate matrix}
In this Section, we discuss how it is possible to reconstruct the full rate matrix $M$ based on the analysis in the previous Secs.~\ref{supp:maximum_likelihood_estimation} and~\ref{supp:extended_error_analysis}.
First, we show a computationally efficient way to directly approximate $M$ from the given jump rates, and afterwards, we show how a global MLE for $M$ can be determined.

\paragraph*{Construction of the full rate matrix.}
So far, we have described how to estimate the diagonal elements $M_{ii}=-\Gamma_i$ of the rate matrix $M$ from the measurement record.
As a subsequent step, we want to argue how it is possible to reconstruct the full matrix including the off-diagonal elements $M_{ji}=\Gamma_{ji}$.
Let us recall that from two measurement sequences
\begin{align}
\label{eq:T_ji_and_T_ki_set}
    \mathcal T_{j|i} = \{\tau_{j|i}^1,\dots,\tau_{j|i}^{N_{j|i}}\},\quad \mathcal T_{k|i} = \{\tau_{k|i}^1,\dots,\tau_{k|i}^{N_{k|i}}\},
\end{align}
we can obtain an estimate for the diagonal value $-\Gamma_i$ of the rate matrix following the estimator~\eqref{eq:hatGamma_full}.
The expectation value of this estimator should be independent of the $k$ and $j$, that is, for the two measurement records, one should obtain the same estimated rate.
Deviations from this behavior would hint at some deviation of the actual dynamics from those predicted by the Markovian master equation.
As we show in Fig.~\ref{fig:cond_rates}, our estimated rates indeed agree with the Markovian model and the predictions from both sets~\eqref{eq:T_ji_and_T_ki_set} agree with each other within their error bars.
Thus, we can estimate the rate $\Gamma_i$ by taking the statistics from both sets~\eqref{eq:T_ji_and_T_ki_set} together, defining
\begin{align}
    \mathcal T_i := \mathcal T_{j|i} \cup \mathcal T_{k|i},
\end{align}
where $k,j$ and $i$ are distinct.
Taking eq.~\eqref{eq:hatGamma_full}, we thus set
\begin{align}
    \widehat\Gamma_i = \frac{N_{j|i}+N_{k|i}}{\sum_{\tau\in\mathcal T_i}(\tau - \widehat\delta_i)},
\end{align}
where $\widehat\delta_i = \min\mathcal T_i$ is the MLE for the detector deadtime.
To recover the off-diagonal elements of $M$, we need to consider $P[j|i]=\Gamma_{ji}/\Gamma_i$ which is the probability that the system jumps from $i$ to $j$, given that a jump occurs.
In the limit of many jumps, we would expect that the fraction $N_{j|i}/(N_{j|i}+N_{k|i})$ approximates $P[j|i]$ which leads us to the estimator for the jump rate $\Gamma_{ji}$,
\begin{align}
\label{eq:HatGamma_ji}
    \widehat\Gamma_{ji} = \frac{N_{j|i}}{N_{k|i}+N_{j|i}} \widehat\Gamma_i.
\end{align}
This estimator fulfills the consistency criterion that $\widehat\Gamma_i = \widehat\Gamma_{ji}+\widehat\Gamma_{ki}$.
The resulting estimate for the rate matrix $\widehat M$ is a valid generator of the stochastic evolution preserving the probability.

\paragraph*{Consistency with full MLE.}
An alternative approach to recover the full rate matrix from the experimentally observed trajectory $(\mathbf s,\mathbf t)$ is to do a maximum likelihood estimation for the full rate matrix $M$.
Since $M$ needs to satisfy the constraint $\sum_{j\in\{0,R,L\}} M_{ji}=0,$ that is $\Gamma_{i}=\Gamma_{ji}+\Gamma_{ki}$ for $i,j,k$ all distinct, the MLE for $M$ consists of $6$ independent variables.
The likelihood function in this case equals the probability of observing the trajectory $(\mathbf s,\mathbf t)$ with a total of $n$ jumps.
Given the state $i$, the probability (density) that the system jumps to state $j$ after a time $\tau$, can be directly calculated from the evolution equation~\eqref{eq:dotp=Mp} to be
\begin{align}
    W[\tau,j|i] &= \Gamma_{ji}e^{-\Gamma_i\tau},
\end{align}
as already used in eq.~\eqref{eq:P[T_ji]_derivation}.
Since we are considering a Markov jump process, the probability of the full trajectory is given by the product,
\begin{align}
\label{eq:P[(i,t)|t]}
    P[\mathbf s,\mathbf t|t] &= e^{-\Gamma_{s_N}(t-t_N)}\left(\prod_{k=0}^{N-1}\Gamma_{s_{k+1}s_k}e^{-\Gamma_{s_k}(t_{k+1}-t_k)}\right) \nonumber \\
    &=\left(\prod_{i\neq j} \Gamma_{ji}^{N_{j|i}}\right)\exp\left(-\sum_{i\in\{0,R,L\}} \Gamma_i\Delta T_i\right),
\end{align}
where $N_{j|i}$ is the number of jumps from $i$ to $j$ and $\Delta T_i$ is the total time the system has hung around in state $i$.
The trajectory probability as defined in eq.~\eqref{eq:P[(i,t)|t]} implicitly conditions on the initial state being $s_0,$ though in the long time limit, this does not affect the MLE for the rates.
We note that the time the system has stayed in state $i$ can be written as
\begin{align}
    \Delta T_i = \sum_{\tau \in \mathcal T_{j|i} \cup \mathcal T_{k|i}}\tau,
\end{align}
that is, the sum over all times $\tau \in \mathcal T_{j|i}\cup \mathcal T_{k|i}\equiv \mathcal T_i$ with $k,j,i$ all distinct.
Recall that we defined $\mathcal T_{j|i}$ in eq.~\eqref{eq:tau_tag_list} as the list of times the system stayed in state $i$ before jumping to $j$.
The log-likelihood function can thus be written as
\begin{align}
    \log P[\mathbf s,\mathbf t|t] = \sum_{i\neq j} N_{j|i}\log\Gamma_{ji} - \sum_{i\in\{0,R,L\}} \Gamma_i \Delta T_i,
\end{align}
and then we can take the derivative with respect to $\Gamma_{ji}$ to determine the MLE.
This yields the equation,
\begin{align}
    0 = \frac{N_{j|i}}{\Gamma_{ji}} - \Delta T_i,
\end{align}
which we can uniquely solve for $\Gamma_{ji}$ to give the MLE
\begin{align}
    \widehat \Gamma_{ji} = \frac{N_{j|i}}{\sum_{\tau \in \mathcal T_i} \tau}.
\end{align}
This is in agreement with our definition in eq.~\eqref{eq:HatGamma_ji}, up to the fact that the former definition may also account for the detector deadtime.

\subsection{\label{supp:theoretical_precision}Theoretical precision}
In this section, we discuss how it is possible to theoretically calculate the precision of a given time estimator $\Theta(t)$ when the rate matrix $M$ is known.
We first start with a general linear counting variable
\begin{align}
\label{eq:Theta_general}
    \Theta(t) = \sum_{i\neq j} \nu_{j|i} N_{j|i}(t),
\end{align}
where $\nu_{j|i}\in\mathbb R$ is an arbitrary but fixed counting weight, and $N_{j|i}(t)$ is the random variable counting how many jumps $i\rightarrow j$ have occurred at time $t$.
Note that $\Theta_{\rm opt}(t)$ as in eq.~\eqref{eq:Theta[i]} and eq.~\eqref{eq:Theta[i]_appendix} is of this form, with the caveat that the former definitions also include a correction of the last state the system has occupied.
That is, we can write
\begin{align}
    \Theta_{\rm opt}(t) &= \sum_{i\in\{0,R,L\}} \frac{n_i(t)}{\Gamma_i} \\
    &= \sum_{i\neq j} \frac{N_{j|i}(t)}{\Gamma_i} + \frac{1}{\Gamma_{i_{m(t)}}},
\end{align}
where $m(t)$ is the number of jumps the system has made until time $t$ and $i_m(t)$ is the state the system is in after the last jump.
The optimal estimator $\Theta_{\rm opt}(t)$ is thus of the general form~\eqref{eq:Theta_general} up to a constant correction due to the last jump.
Similarly, looking at the net number of charges that have been transported across the DQD, the estimator can be written as
\begin{align}
    \Theta_{\rm net}(t) &= \frac{1}{\nu}N(t) \\
    &= \frac{1}{\nu}\left(N_{0|R}(t) -N_{R|0}(t) \pm 1\right),
\end{align}
where the correction $\pm 1$ comes from the difference of the net charges $N(t)$ that have tunneled across the DQD, and the net charges $N_{0|R}(t) -N_{R|0}(t)$ that have transported from the right dot to the right reservoir.
The reason the difference in these two numbers is at most $1$, is the fact that the DQD in the regime it is used can at most be occupied with a single charge.
With charge conservation, we have that
\begin{align}
    N_{L|0} - N_{0|L} = N_{0|R} -N_{R|0} + (1-\delta_{s0}),
\end{align}
where $N_{L|0} - N_{0|L}$ is the net number of charges that have tunneled from the left reservoir into the DQD.
On the other hand, $N_{0|R} -N_{R|0}$ equals the net number of charges that have tunneled from the DQD out into the right reservoir, and $1-\delta_{s0}$ is the number of charges in the DQD.
That is, if the DQD is empty the state is $s=0$ thus there are no charges in the DQD.
If a charge is in the left or right, $s=L,R$ there is one charge in the DQD.
In summary, the net charges that have tunneled fully across the DQD equal the net charges that have gone to the right reservoir $\pm 1$ coming from the correction due to a charge possibly in the DQD.
In the long time limit, this $O(1)$ correction to $\Theta$ does not affect the precision, because the first two cumulants $\E[\Theta(t)]$ and $\Var[\Theta(t)]$ are only affected to subleading order.

We thus show how the asymptotic precision $\mathcal S$ can be calculated for the estimator $\Theta(t)$ of the form~\eqref{eq:Theta_general}, from which it follows how the precision can be calculated for the two estimators $\Theta_{\rm opt}(t)$ and $\Theta_{\rm net}(t)$.
Making use of full counting statistics, we can determine the statistics of $\Theta(t)$ by introducing a full-counting field $\chi$~\cite{Schaller2014}.
The full-counting field can be treated as a formal, real variable used to define the tilted rate matrix $M(\chi)$ defined by the elements
\begin{align}
    M_{ji}(\chi) = M_{ji} e^{i \nu_{ji}\chi},
\end{align}
for $j\neq i$ and $M_{ii}(\chi)=M_{ii}$ is the same as in the non-tilted case.
When we thus replace the usual evolution equation~\eqref{eq:dotp=Mp} for our system with $\mathbf {\dot p}(t,\chi) = M(\chi)\mathbf p(t,\chi)$, the solution $\mathbf p(t,\chi)$ encodes the system's full jump statistics,
\begin{align}
    \mathbf p(t,\chi) &= \sum_{\{N_{j|i}\}} \mathbf p_{\{N_{j|i}\}}(t) e^{i\chi \sum_{i\neq j}\nu_{ji}N_{j|i}} \\
    &= \sum_{\Theta}\mathbf p_{\Theta}(t) e^{i\chi\Theta}.
\end{align}
The norm $\|\mathbf p_\Theta(t)\|_1$ equals the probability that at time $t$, the estimator takes the value $\Theta$.
Thus, $\mu(t,\chi)=\|p(t,\chi)\|_1$ is the moment generating function and we can write $\E[\Theta(t)] = -i\partial_\chi \mu(t,\chi)|_{\chi=0}$ for the expectation value as well as $\Var[\Theta(t)]=-\partial_\chi^2  \mu(t,\chi)|_{\chi=0} - \E[\Theta(t)]^2$ for the variance.
Further following the derivation in~\cite{Schaller2014}, the cumulant generating function $c(t,\chi)=\log\mu(t,\chi)$ can be obtained directly from the dominant eigenvalue of $M(\chi)$ if $M$ admits a unique steady-state.
It is given by,
\begin{align}
    c(t,\chi) = \lambda_*(\chi) t + O(1),
\end{align}
in the limit $t\rightarrow\infty$ and where $\lambda_*(\chi)$ is the eigenvalue of $M(\chi)$ with the largest real part.
We can then directly calculate
\begin{align}
    \frac{\E[\Theta(t)]}{t} = -i\partial_\chi \lambda_*(\chi)\big|_{\chi=0} + o(1),
\end{align}
and
\begin{align}
    \frac{\Var[\Theta(t)]}{t} = -\partial_\chi^2 \lambda_*(\chi)\big|_{\chi=0} + o(1),
\end{align}
both in the limit $t\rightarrow\infty$.
In summary, we can calculate
\begin{align}
\label{eq:S_th}
    \mathcal S = -\frac{\left(\partial_\chi \lambda_*(\chi)\right)^2}{\partial_\chi^2 \lambda_*(\chi)}\bigg|_{\chi=0},
\end{align}
to obtain the precision directly from the dominant eigenvalue $\lambda_*(\chi)$ of $M(\chi)$.
This concludes how precision $\mathcal S$ can be obtained theoretically from the rate matrix $M$.

\paragraph*{Error analysis.}
The precision as theoretically calculated in eq.~\eqref{eq:S_th} is a function of the underlying rate matrix $M$.
When we estimate $M$, there is an associated error with the underlying rates, which propagates to become an error of the theoretically calculated precision.
In its general form, the standard error of $\mathcal S$ can be estimated using Gauss error propagation
\begin{align}
    \widehat\sigma_{\mathcal S}^2 = \sum_{i\neq j} \left(\frac{\partial \mathcal S}{\partial M_{ji}}\bigg|_{M_{ji}=\widehat\Gamma_{ji}}\right)^2 \widehat\delta{\Gamma_{ji}}^2.
\end{align}
The error bars shown in Fig.~\ref{fig:panel2}(a) of the main text are given by $\widehat\sigma_{\mathcal S}$.
For the optimal estimator, the precision has been calculated analytically in Ref.~\cite{Prech2024},
\begin{align}
    \mathcal S_{\rm opt} = \left(\sum_{i\in\{0,R,L\}} \frac{p_i^{\rm s.s.}}{\Gamma_i}\right)^{-1},
\end{align}
which allows for a more direct computation of the estimated standard error.

\begin{figure*}
    \centering
    \includegraphics[width=\linewidth]{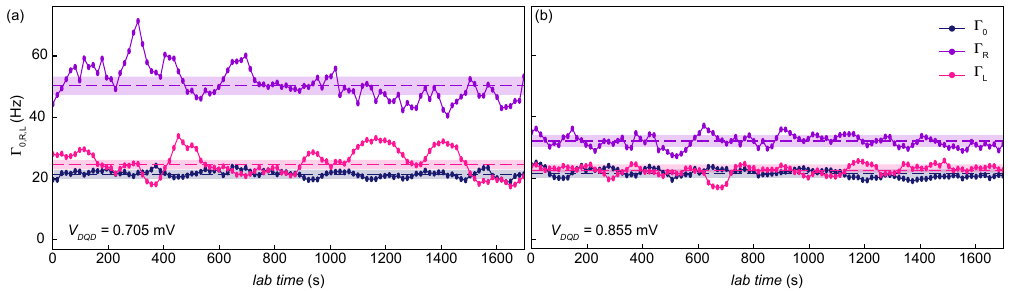}
    \caption{Comparison of the diagonal elements of the rate matrix estimate $M$ for the two DQD bias settings $V_{\rm DQD}=0.705\,{\rm mV}$ (top) and $V_{\rm DQD}=0.855\,{\rm mV}$ (bottom).
    The data shown is obtained through the dc measurement for maximal sensor bias $V_{\rm cs}=0.345\,{\rm mV}$.
    The dashed horizontal lines are the rates estimates as obtained for the entire time trace, together with the standard error estimates given as the transparent band (see Sec.~\ref{supp:extended_error_analysis} for details).
    The dotted line are the estimates for the rate matrix using only a subset of the time trace of length $\Delta\approx 80\sec$ as described in Sec.~\ref{supp:rate_drifts}.
    The $x$-axis value of each point is the starting point of the interval.
    We see that for some settings like e.g. the $V_{\rm DQD}=0.705\,{\rm mV}$, the rates deviate from the value obtained from the entire time trace beyond the standard error.
    In comparison, for $V_{\rm DQD}=0.855\,{\rm mV}$, the instantaneous rates are more stable.    
    This indicates that the Gaussian white noise model is insufficient in capturing drifts that affect the device's parameters in some settings.
    This deviation from a model with fixed rate matrix may also lead to the difference between the theoretical precision estimates and the empirical precision estimate shown in the third data point from the right in Fig.~\ref{fig:panel2} of the main text.}
    \label{fig:rate_stability}
\end{figure*}

\subsection{\label{supp:rate_drifts}Rate drifts}
In this section, we assess the stability of the transition rates of the DQD during the $\approx 30\min$ time trace, as also visualized in Fig.~\ref{fig:rate_stability}.
In the previous Secs.~\ref{supp:maximum_likelihood_estimation} and~\ref{supp:extended_error_analysis}, we determined the maximum likelihood estimator for the matrix elements of the rate matrix $M$ describing the DQD evolution.
There, we assumed constant rates for the duration of the experiment.
For the error analysis, however, we allowed Gaussian random noise to affect the rates.
The question we now investigate is how well these assumptions are satisfied within the experiment.

To this end, we consider sub-samples of a duration of approximately $\Delta \sim 80\sec$ of the full trace $S(t)$.
These are defined on the time-windows $\mathcal T_0,\mathcal T_1,\dots$ as follows
\begin{align}
    \mathcal T_k = [kq\Delta,(kq+1)\Delta],
\end{align}
where we have chosen $q=0.2$.
In other words, the time windows $\mathcal T_k$ are intervals of width $\Delta$ that are each shifted by a fraction of their length $q\Delta$ from each other.
By considering the subtraces of the full time trace
\begin{align}
    \{S(t)\}_{t\in\mathcal T_k},
\end{align}
we can obtain a MLE estimate for the rate matrix $M$ for each individual slice $\mathcal T_k$.
In Fig.~\ref{fig:rate_stability}, we compare two examplary estimates for the diagonal elements of the rate matrix.
Because the data sets for the times $\mathcal T_k$ are shorter than the full data set, the intrinsic errors coming from the stochastic nature of the jump processes lead to a bigger uncertainty for the estimates derived from the reduced sets in comparison to the full data set.
To ensure enough data points are available, the time window needs to be much larger than the time scale of the jump process, $\Delta\gg 1/\Gamma$.
This is to guarantee that a significant sample of jump times lies within the given time window to estimate the underlying rates.
For the case of the data shown in Fig.~\ref{fig:rate_stability}, we have $\Gamma \sim 20\,{\rm Hz} -  60\,{\rm Hz}$, and $\Delta \sim 80\sec,$ this assumption is satisfied.

\section{\label{supp:general_clock_theory}General quantum clock theory}
In this section, we provide additional detail describing quantum clocks in a more general setting beyond the three-state system considered as part of this work.

\paragraph*{Setup.}
General quantum clocks can be described with a quantum master equation given the interactions with the environments are sufficiently well-behaved for the Markovian assumption to be satisfied.
The clockwork's state is given by some quantum state $\rho$ evolving according to $\dot\rho = \mathcal L \rho$ with Liouville superoperator $\mathcal L= -i[H,\,\cdot\,] + \sum_k \mathcal L_k$.
The superoperator encodes the clockwork's Hamiltonian $H$ and interactions with the thermal environments $E_k$ which can be split into two contributions $\mathcal L_k=\mathcal L_k^+ + \mathcal L_k^-$.
One process $\mathcal L_k^+$ describes a charge exiting the system through interactions with environment $k$ and $\mathcal L_k^-$ describes the reverse process, i.e. a charge entering the system.
Formally, these processes can be described with jump operators $J_k^\pm$ such that $\mathcal L_k^{\pm} = J_k^\pm \,\cdot \, J_k^{\pm\dagger}-\tfrac{1}{2}\{J_k^{\pm\dagger}J_k^\pm,\,\cdot\,\}$.
The ratio of probability for a charge entering the environment versus leaving is then related to the entropy production $s_k$ in the environment via $\|J_k^+\|^2 /\|J_k^-\|^2 = e^{s_k}$.

\paragraph*{Time estimation on the quantum level.}
As time progresses, the clockwork exchanges charges across the thermal bias between the environments, for example a temperature gradient or a difference in chemical potentials.
While the exchange of charges can be understood as the \textit{ticks} of the clock, the net number of charges $N_k(t)=N_k^+(t)-N_k^-(t)$ that have been exchanged with some environment $E_k$ define the clock's reading~\cite{Silva2023}.
Here $N_k^+$ counts the number of jumps described by the operator $J_k^+$ (charges leaving the system) and $N_k^-$ counts the corresponding reverse processes described by $J_k^-$.
At the level of the information encoded in the thermal environments, the clock's time estimate can in principle be any function of the $N_k(t)$'s.
One class of estimators is formed by the linear combinations
\begin{align}
\label{eq:Theta_antisymm}
    \Theta(t) = \sum_k \nu_kN_k(t).
\end{align}
The weights $\nu_k$ are the rates with which the different counting variables are counted towards the time estimate, ideally chosen such that the estimator is unbiased, $\E[\Theta(t)]=t+O(1)$.
Estimators of this form define an antisymmetric current because they are antisymmetric under swapping forward and backward processes $N_k^+\leftrightarrow N_k^-$.
For the clock's reading to monotonously change in time, the probability for a charge leaving the system, 
\begin{align}
    dN_k^+ = \tr[ J_k^+\rho J_k^{+\dagger}]dt,
\end{align}
must differ from that of a charge entering the system,
\begin{align}
    dN_k^- = \tr[ J_k^-\rho J_k^{-\dagger}]dt,    
\end{align}
otherwise $\E[N_k(t)]$ is constant and the clock stands still on average.
This condition in particular implies non-zero entropy production 
\begin{align}
    \Sigma_{\rm cw}(t)=\sum_k\int_0^t d N_k s_k>0   
\end{align}
due to the second law~\cite{Landi2021}.
This corresponds to the clockwork cost also shown in Fig.~\ref{fig:panel}(a).
The clockwork entropy cost can be related to the cost per tick used in eq.~\eqref{eq:SigmaTick} of the main text $\Sigma_{\rm tick}$, by writing $\nu\Sigma_{\rm tick}=\dot\Sigma_{\rm cw}$, where $\nu$ is the tick rate.
The precision of any antisymmetric estimator of the form~\eqref{eq:Theta_antisymm}, i.e.~only linearly depending on the net number of charge exchanges, is usually bounded by the clockwork dissipation.
In case of the present experiment, where the dynamics are fully incoherent, the thermodynamic uncertainty relations~\cite{Barato2015,Gingrich2016,Pietzonka2017,Horowitz2020} provide one such bound,
\begin{align}
\label{eq:S_TUR}
    \mathcal S \leq \frac{\dot\Sigma_{\rm cw}}{2},
\end{align}
where $\mathcal S$ is defined as in eq.~\eqref{eq:S}.
Thus, when the clockwork is at equilibrium ($\dot\Sigma_{\rm cw} = 0$), the clock at the quantum level stands still on average ($\mathcal S = 0$, implying that the tick rate $\nu = 0$, although the diffusion $D>0$, implying a growing uncertainty in the number of ticks).

For the more general quantum setting, a universal bound of similar form as~\eqref{eq:S_TUR} does not exist and is still actively under research~\cite{Guarnieri2019, Hasegawa2020,Hasegawa2021,Vu2022,Moreira2024}. 
However, even in this case, one can argue that the precision is $0$ if the clockwork dissipation is $0$.
Consider a quantum-mechanical clockwork coupled to a number of thermal environments.
The only non-trivial way the clockwork dissipation is zero in the long-time limit is if the environments are all in thermal equilibrium with each other, i.e., at the same temperature. 
In this case the clock reaches equilibrium at the same temperature, and every transition in the clockwork that is coupled to a bath is in equilibrium, therefore there is no net charge transferred to or from the bath.
Hence any anti-symmetric current observable constructed to count the number of ticks will be zero in the long-time limit, leading to zero precision.

\paragraph*{Time estimation with a classical record.}
However, when creating a classical record of the quantum clockwork's evolution like in the present experiment, not only the instantaneous values $N_k(t)$ are available, but the full history, though without the time tags $t$.
This is to guarantee that no external clock is used to record the trajectory, and thus accidentally serves as a resource for the time estimate.
The full trajectory is a sequence of the form $\mathbf{\Gamma}(t) = (k_1^{\sigma_1},k_2^{\sigma_2},\dots,k_{m(t)}^{\sigma_{m(t)}})$, where $m(t)$ is the total number of jumps observed up to the instantaneous time $t$, and $k_1^{\sigma_1},k_2^{\sigma_2},\dots$ is the ordered sequence of jumps.
That is, each index $k_i^{\sigma_i}$ in the trajectory corresponds to one of the jump channels $k$ of the system and $\sigma_i=\pm$ encodes the information of whether it was a forward jump generated by $J_k^+$ or a backward jump by $J_k^-$.
In principle, a time estimator based on the full trajectory can be any function $\Theta[\mathbf\Gamma]$ which is (ideally) unbiased in the long time limit $\E[\Theta[\mathbf\Gamma(t)]]=t+O(1)$.
Which estimator is optimal in a general open quantum system's setting, however, is still an open question, albeit some partial answers for the case of renewal-processes have been given recently in Ref.~\cite{Macieszczak2024}.
In general, such estimators do not satisfy uncertainty relations akin to the inequality~\eqref{eq:S_TUR} with the entropy production, and in fact, they can arbitrarily `violate' such bounds since transitions can be observed even at equilibrium where $\dot\Sigma_{\rm cw}=0$.

In the case where the clockwork dynamics are fully dissipative i.e., the evolution is governed by a classical master equation with vanishing Hamiltonian $H=0$, an optimal choice for the time estimator can be explicitly constructed~\cite{Prech2024}.
Here, optimality refers to the minimization of the variance $\Var[\Theta(t)]$ given the constraint that the estimator is unbiased.
The optimal estimator which we have considered for the DQD experiment follows from Ref.~\cite{Prech2024}.
For the experiment, a jump trajectory $\mathbf\Gamma(t)$ always uniquely defines a state trajectory $\mathbf s(t)$ as defined in the main text,
\begin{align}
\label{eq:Gamma_S_equiv}
    \mathbf\Gamma(t)= \left(k_1^{\sigma_1},\dots,k_{m(t)}^{\sigma_{m(t)}}\right) \simeq \mathbf s(t) = \left(s_0,s_1,\dots,s_{m(t)}\right),
\end{align}
because every jump uniquely characterizes a state before the jump and one after.
We recall that $s_n$ is the state the system occupies after the $n$th state; $s_0$ is the initial state.
For more general systems where a jump may not uniquely define the post-jump state, there may not exist a one-to-one relationship between $\mathbf\Gamma(t)$ and $\mathbf s(t)$.
The optimal estimator can thus be written as a function of the recorded trajectory due to the relation~\eqref{eq:Gamma_S_equiv},
\begin{align}
    \Theta_{\rm opt}[\mathbf \Gamma (t)] = \sum_{n=0}^{m(t)}\frac{1}{\Gamma_{s_n}},
\end{align}
as also defined in eq.~\eqref{eq:Theta[i]} in the main text, or eq.~\eqref{eq:Theta[i]_appendix} in Sec.~\ref{supp:empirical_precision}
As also the experimental results in Fig.~\ref{fig:panel2} show, the precision $\mathcal S$ of the optimal estimator is non-zero even when the clockwork is at equilibrium, because the full trajectory information is used in the estimator, showing that $\mathcal S>0$ is compatible with $\dot\Sigma_{\rm cw}=0$.

The ultimate reason why the thermodynamic uncertainty relations break down for estimators based on the full trajectory is the fact that formally, creating the trajectory record $\mathbf \Gamma(t)$ is a fully irreversible process.
That is, if for example a jump $J_k^+$ occurs in the quantum clockwork, this triggers a transition in the tick record,
\begin{align}
    \{\dots,k_{i}^{\sigma_{i}}\} \rightarrow \{\dots,k_i^{\sigma_i},k^+\}.
\end{align}
While there exists the reverse process $J_k^-$ on the level of the quantum clockwork, the reverse process on the tick record does formally not exist, i.e., the jump $J_k^-$ leads to a transition in the tick record
\begin{align}
    \{\dots,k_{i}^{\sigma_{i}},k^+\} \rightarrow \{\dots,k_{i}^{\sigma_{i}},k^+,k^-\},
\end{align}
and not to $\{\dots k_{i}^{\sigma_{i}}\}$.
This is because the tick record includes, by assumption, the full history of the jumps.
Formally, however, the absence of the reverse process in the jump memory implies an infinite entropy production due to detailed balance.

In practice, the irreversibility comes from the measurement necessary to create the jump record~\cite{Granger2011,Deffner2016,Debarba2019,Guryanova2020} as well as the redundant encoding of the trajectory information in macroscopic systems.
Realistically, there is never infinite entropy production, and thus, the full irreversibility of the recording is an idealization.
For example, in the present DQD experiment, the entropy dissipated by the measurement per jump was approximately between $10^9\,k_B$ and $10^{11}\,k_B$, several orders of magnitude larger than the dissipation within the clockwork.
For all practical purposes the classical record of the trajectory can thus be considered irreversible and permanent.
\end{appendices}

\end{document}